\documentclass[12pt]{article}
\usepackage[dvips]{epsfig}
\usepackage{amsmath,amssymb,amsfonts,enumerate,bbm}
\usepackage{caption}
\usepackage{url}
\usepackage{ifthen}

\usepackage{graphicx}
\usepackage{fullpage}
\usepackage{ifthen}
\usepackage{ifpdf}

\usepackage{subfigure}

\usepackage{color,soul}
\usepackage[dvips]{graphicx}

\newtheorem{theorem}{Theorem}[section]
\newtheorem{lemma}[theorem]{Lemma}
\newtheorem{observation}[theorem]{Observation}
\newtheorem{corollary}[theorem]{Corollary}

\def\Proof{\par\noindent{\bf Proof:~}}

\long\def\commabs #1\commabsend{}
\long\def\commful #1\commfulend{#1}

\newcommand{\R}{\mathbb{R}}

\def\CancelTree2D{\mbox{\sf CancelTree2D}}

\def\HDS{\mbox{\sf BuildHDS}}
\def\LabelArrangment{\mbox{\sf LabelArrangment}}
\def\LabelCell{\mbox{\sf LabelCell}}
\def\ExtractNCO{\mbox{\sf ExtractNCO}}

\newcommand{\Station}[0]{\mathit{s}}

\newcommand{\cA}{\mathcal{A}}
\newcommand{\Power}[0]{\psi}

\newcommand{\Noise}[0]{\mathit{N}}
\newcommand{\Reals}[0]{\mathbb{R}}
\newcommand{\Integers}[0]{\mathbb{Z}}

\newcommand{\Area}[0]{\mathrm{area}}

\newcommand{\DataStructure}[0]{\ensuremath{\mathtt{DS}}}

\newcommand{\Energy}[0]{\mathrm{E}}

\newcommand{\ReceptionZone}[0]{\mathcal{H}}
\newcommand{\NZones}[0]{\mathcal{\tau}}

\newcommand{\SINR}[0]{\mathrm{SINR}}

\newcommand{\Ball}[0]{\mathit{B}}

\newcommand{\dist}[1]{\mathrm{dist} ({#1})}
\newcommand{\maxdist}[1]{\mathrm{maxdist} ({#1})}
\newcommand{\mindist}[1]{\mathrm{mindist} ({#1})}

\newcommand{\NVorZonesArrangment}[0]{\mathcal{\tau}^{A}}
\newcommand{\CompactnessParameter}[0]{\mathcal{CP}}

\newcommand{\HH}[0]{\mathcal{H}}
\newcommand{\HCON}[0]{\mathcal{H}}
\newcommand{\HORDER}[1]{\mathcal{H}(\overrightarrow{#1})}
\newcommand{\Arrangment}[0]{Ar}

\newcommand{\vor}[0]{\mbox{\sc Vor}}

\newcommand{\vororder}[1]{\mbox{\sc Vor}(\overrightarrow{#1})}

\def\orderedSi{\overrightarrow{S_i}}
\def\orderedSj{\overrightarrow{S_j}}
\def\orderedSp{\overrightarrow{S_p}}

\def\LAST{{\sf Last}}
\def\CO{{\cal CO}}
\newcommand{\NCO}[0]{{\cal NCO}}

\newcommand{\HORDERQ}[1]{\mathcal{H}^?\left(\overrightarrow{#1}\right)}
\def\Intersect{\mbox{\sf Intersect}}
\newcommand{\Ceil[1]}{\ensuremath{\left\lceil#1\right\rceil}}
\def\blackslug{\hbox{\hskip 1pt \vrule width 4pt height 8pt
    depth 1.5pt \hskip 1pt}}
\def\QED{\quad\blackslug\lower 8.5pt\null\par}

\def\keywords{\vspace{.5em}
{\textit{Keywords}:\,\relax%
}}

\begin{document}



\title{SINR Diagram with Interference Cancellation}
\author{
Chen Avin
\thanks{
\mbox{Ben Gurion University, Beer-Sheva, Israel.
}
\mbox{\tt \{avin,coasaf,zvilo\}@cse.bgu.ac.il, yoram.haddad@gmail.com}}
\and
Asaf Cohen $^*$
\and
Yoram Haddad $^*$
\thanks{Jerusalem College of Technology,Jerusalem, Israel}
\and
Erez Kantor
\thanks{
\mbox{The Technion, Haifa, Israel.
{\tt erez.kantor@gmail.com}}}
\and
Zvi Lotker $^{*\P}$
\and
Merav Parter $^{\S\P}$
\and
David Peleg
\thanks{
The Weizmann Institute of Science, Rehovot, Israel.
{\tt \{merav.parter,david.peleg\}@ weizmann.ac.il}.}
\thanks{Supported in part by a grant of the Israel Science Foundation and
the I-CORE program of the Israel PBC and ISF (grant 4/11).}
\thanks{Parts of this work appeared at the ACM-SIAM Symposium on Discrete Algorithms (SODA), Kyoto, Japan, 2012.}
}

\maketitle

\setcounter{page}{1}

\begin{abstract} \small\baselineskip=9pt
In this paper we study the reception zones of a wireless network in the SINR model with receivers that employ \emph{interference cancellation} (IC), a technique that allows
a receiver to decode interfering signals, and \emph{cancel} them from the received signal in order to decode its intended message. We first derive some important topological properties of the diagram describing the reception zones
and their connections to \emph{high-order Voronoi diagrams} and other related geometric objects.
We then discuss the computational issues that arise when seeking an efficient description of  the zones.
Our main fundamental result states that although potentially there are
exponentially many possible cancellation orderings (and consequently reception cells), in fact there are much fewer nonempty such cells.
We prove a (tight) linear bound on the number of cells and provide a polynomial time algorithm to describe the diagram.
Moreover, we introduce a novel measure, referred to as the \emph{Compactness Parameter}, which influences the tightness
of our bounds.
We then utilize the properties established for reception diagrams to devise a logarithmic time algorithm for answering \emph{point-location} queries
for networks with IC.
\end{abstract}

\keywords{Interference cancellation, SINR, Voronoi diagram.}

\section{Introduction}
\subsection{Background and Motivation}
Today, wireless communication is embedded in our daily lives, with an ever-growing use of cellular, satellite and sensor networks.
The major advantage of wireless communication, namely, the broadcast nature of the medium, also creates its greatest obstacle, namely, interference. When a station has to decode a message (i.e., a signal) sent from a transmitter, it must cope with all other (legitimate) simultaneous neighboring transmissions.

Roughly speaking, two basic approaches to handling interference dominated  the research community for many years \cite{etkin2008gaussian}. One approache is \emph{orthogonalization}. By using, e.g., time-division (TDMA) or frequency division (FDMA), the degrees of freedom in the channel can be divided among the participating transmitters. This generates an independent channel for each transmitter. The second approach is to treat the interference as \emph{noise}. Taking this view, the interference, together with the ambient (or background) noise, disrupt the signal reception and decoding abilities. For the signal to be safely decoded, the {\em Signal to Interference \& Noise Ratio (SINR)} must be large enough.

Due to the increasingly large number of users, the \emph{achievable rate} or \emph{utilization} of wireless networks has become the bottleneck of the communication.
Consequently, the capacity of wireless networks, i.e., the maximum \emph{achievable rate} by which stations can communicate reliably, has received increasing attention in recent years \cite{AD09,ALP09,GHW09,GuKu00,HW09,Mo07,OLT07}.
One of the main challenges for wireless network designers is to increase this rate and try to fully utilize the capacity of the network.
In a sense, both of the aforementioned approaches treat interference in wireless communication  as a foe, and try to either avoid it or overcome it.
However, modern coding techniques suggest the ability to jointly decode several signals simultaneously, achieving a higher total capacity (see, e.g., Chapter 14 of \cite{Cov_Thom91-short}).
This paper focuses on a relatively recent and promising method for such joint decoding called \emph{interference cancellation (IC)} \cite{Andrews2005}.

The basic idea of interference cancellation, and in particular {\em successive interference cancellation (SIC)}, is quite simple. Consider a situation where a station receives a ``combined" transmission composed of several interfering signal that were transmitted simultaneously.
The station is interested in decoding one of those signals, the ``intended'' signal. However, it might not be the dominant signal in the combined transmission received by the station.
The station can attempt the following technique for retrieving its intended signal.
First, the strongest interfering signal is detected and decoded. Once decoded, this signal can then be subtracted (``canceled") from the received (combined) signal. Subsequently, the next strongest interfering signal can be detected and decoded from the now ``cleaner" combined signal, and so on.  Optimally, this process continues until all stronger interferences are cancelled and we are left with the desired intended signal, which can now be decoded.
This successive process seems prone to error propagation. Nevertheless, note that if the SINR while decoding a message at a given iteration is high enough (above a threshold $\beta$), then the process ensures correct decoding at that stage (at high probability), and using simple union bound, correct decoding at all stages.
SIC is similar in spirit to several well known algorithms like the Gram-Schmidt process \cite{trefethen1997numerical}, solving triangular systems of linear equations, and fountain codes \cite{byers1998digital-short}. 
It should be noted that without using IC, every station can decode at most one transmitter (i.e., the strongest signal it receives). In contrast, with IC, stations can potentially decode more transmitters, or expressed dually, every transmitter can potentially reach more receivers. This clearly increases the utilization of the network.

The SINR model and interference cancellation are fairly well-studied from an information-theoretic point of view. The SINR model was used in the analysis of network capacity and throughput, e.g., \cite{GuKu00} and the many papers which followed. Scheduling under the SINR model was discussed in \cite{pantelidou2010cross}, while random access techniques were given in \cite{louie2011open}. In \cite{weber2005transmission}, the authors considered a stochastic SINR model, where the focus was on the outage probability - the probability that a receiver's SINR is below a threshold $\beta$. In \cite{mahdavi2010characterization}, the sets of possible SINR values subject to linear power constraints were characterized. The zero-outage region, the region achievable regardless of the channel realization, was also considered. \cite{mahdavi2010characterization} also considered the problem of removing a subset of the users in order do achieve certain SINR demands. A suboptimal algorithm maximizing the number of active users was suggested.

Interference cancellation is the optimal strategy in several scenarios, such as strong interference \cite{sato1981capacity,costa1987capacity-short}, corner points of a multiple access channel \cite[Chapter 14]{Cov_Thom91-short}, and spread spectrum communication (CDMA) \cite{viterbi1990very}, and it constitutes a key building block in the best known bounds for the capacity of the interference channel \cite{etkin2008gaussian}. \cite{weber2007transmission} studied the transmission capacity of wireless ad-hoc networks under successive interference cancellation. Although considering SIC under the SINR regime, \cite{weber2007transmission} focused on the capacity and outage probability, rather than the geometric and algorithmic aspects of the reception zones.
Thus, to the best of our knowledge, little is known about the structure and the properties of the reception zones under interference cancellation (namely, the areas in which transmitters can be decoded), as well as algorithmic issues for large wireless networks.

In this paper, we initiate the study of the topological properties of the reception zones in the context of the IC
setting, discuss the computational issues arising when trying to compute these reception zones or answer queries regarding specific points, and devise polynomial-time algorithms to address these problems.
This is done by extending the notion of \emph{SINR diagrams} \cite{Avin2009PODC+full} to the setting of stations that can apply successive interference cancellation.
The SINR diagram of a wireless network of $n$ transmitters $s_1, s_2, \dots s_n$ partitions the
plane into reception zones $\HH(\Station_1), \HH(\Station_2), \dots \HH(\Station_n)$, one per station, and a complementary
region of the plane where no station can be decoded, denoted $\HH(\emptyset)$.
In \cite{Avin2009PODC+full}, SINR diagrams have been studied for the specific case where all stations use the same transmission power, i.e., \emph{uniform power}. It is shown therein that the reception zones have some ``nice" properties, like being convex (hence connected) and ``fat" (as defined later on).
In  \cite{KLPP2011STOC} it was established that for a \emph{nonuniform power} setting, the reception zones are not necessarily connected, but are (perhaps surprisingly) hyperbolically convex in a space of dimension higher by one than the network's dimension. Turning to the stochastic setting, the relation between stochastic SINR diagram (formed by modeling the SINR as a marked point process) and classical stochastic geometry models such as Poisson–Voronoi tessellations, has been studied extensively. See \cite{Baccelli09} for a detailed analysis, results and applications of this approach.


\begin{figure*}
\begin{center}
(a) \includegraphics[scale=0.45]{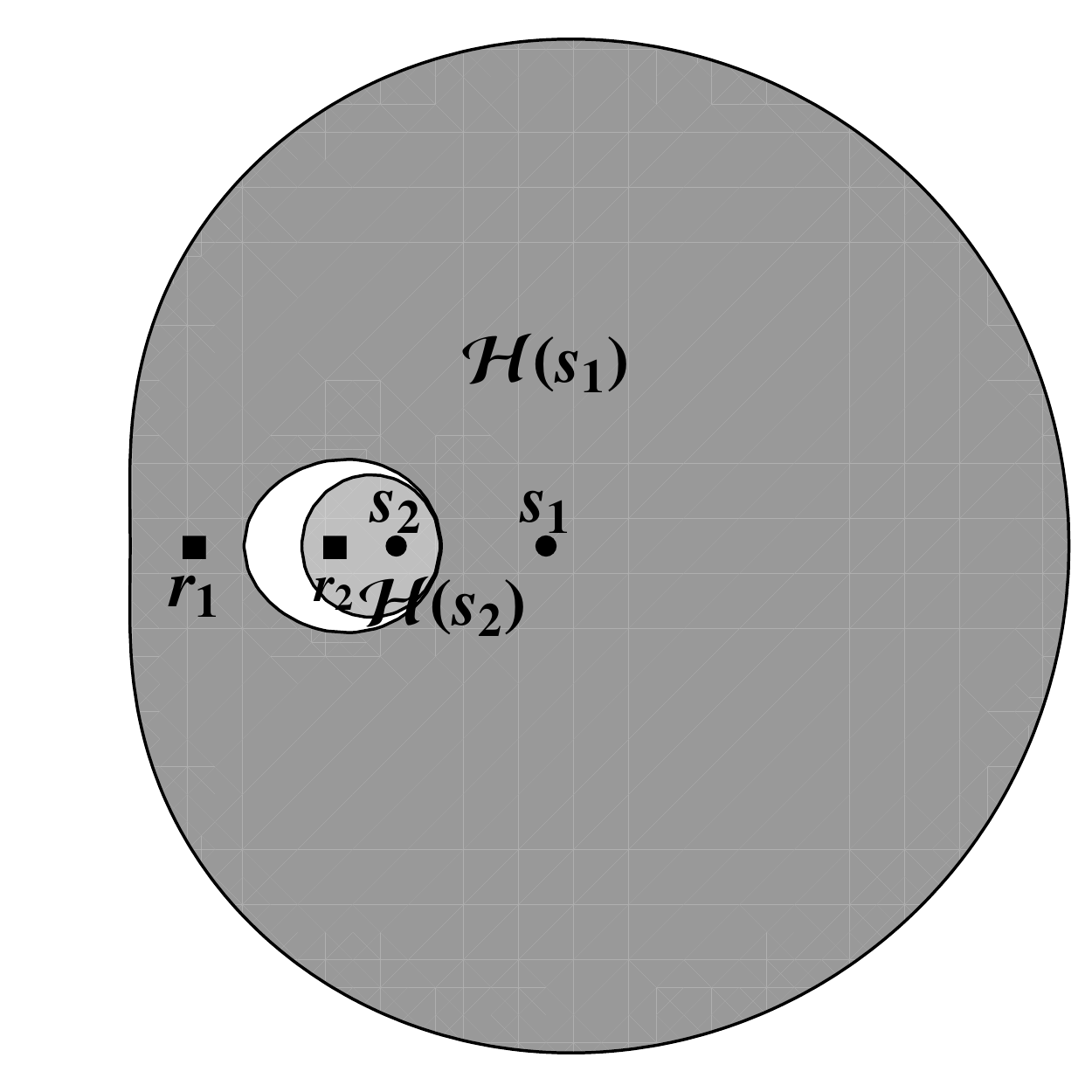}
\hfill
(b) \includegraphics[scale=0.45]{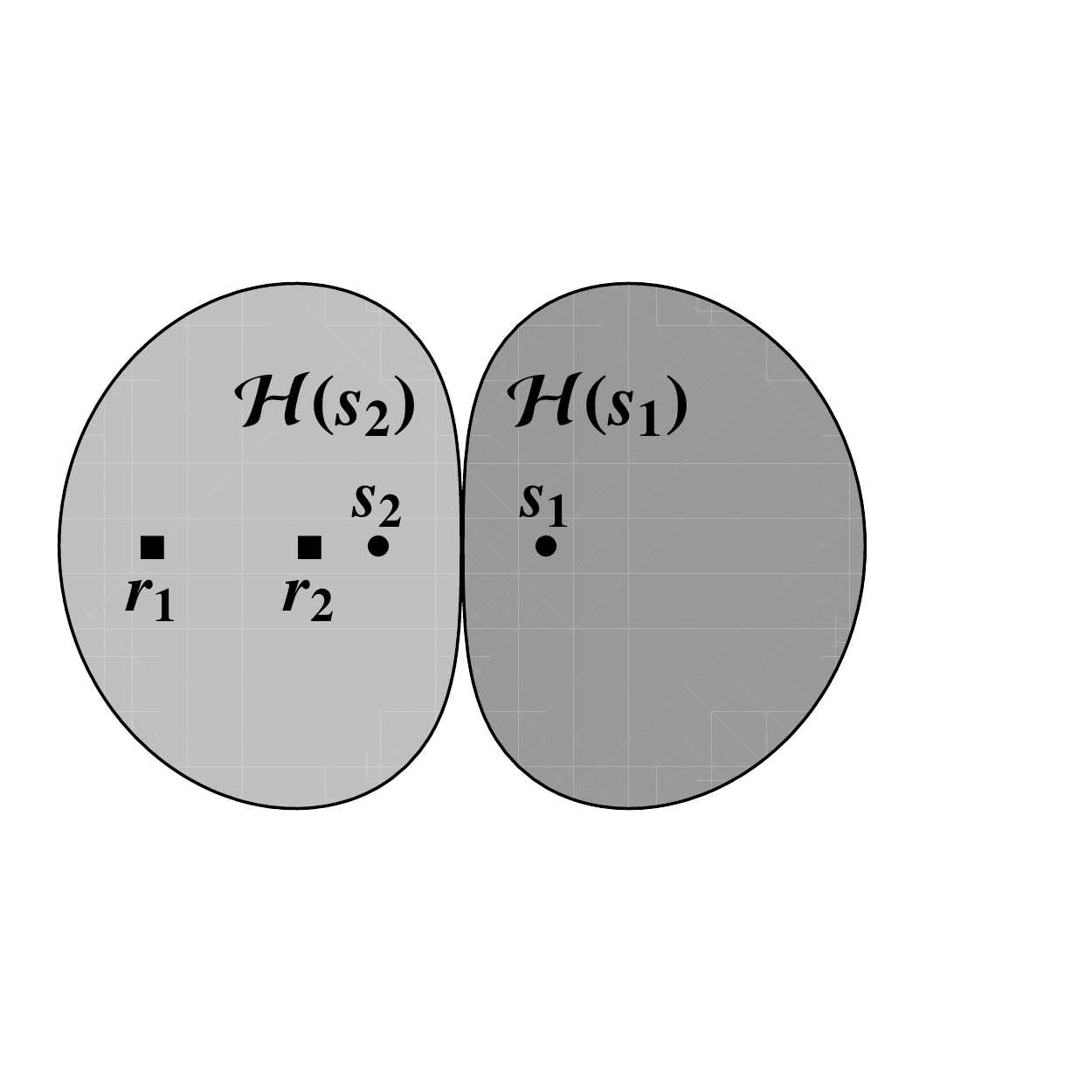}
\hfill
\includegraphics[scale=0.45]{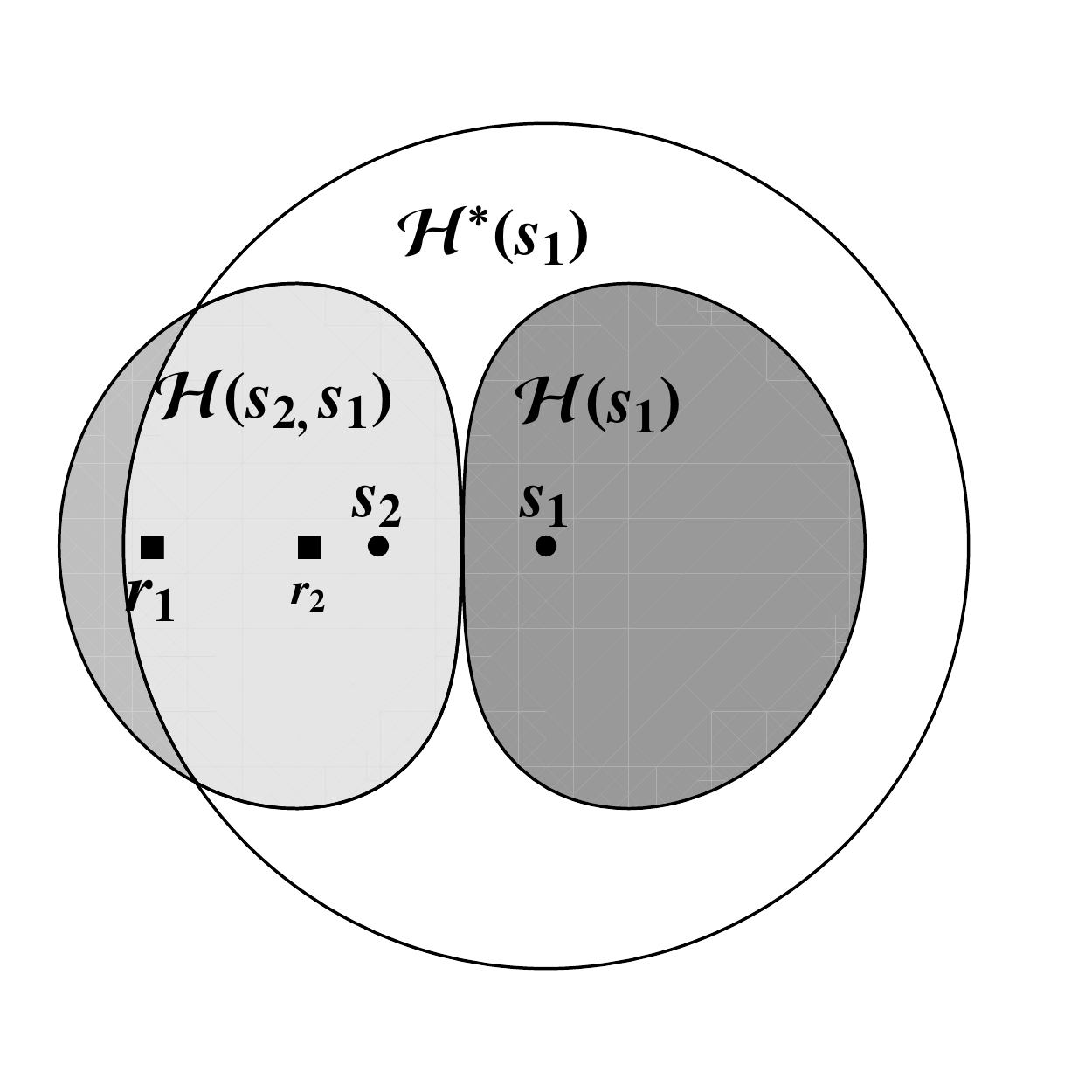}\\(c)
\end{center}
\caption{(a) \emph{Nonuniform} power assignment where for $i=1,2$, receiver $r_i$ is in the reception zone $\HH(\Station_i)$ of transmitter $s_i$.
(b) In every \emph{uniform} power assignment, $r_1$ is not in the reception zone of $s_1$.
(c) In a \emph{uniform} power setting but with interference cancellation, station $r_1$ may be in the reception zone $\HH(\Station_2,\Station_1)$ of transmitter $s_1$, i.e., it could decode transmitter $s_1$ after canceling transmitter $s_2$.}
\label{fig:intro}
\end{figure*}
%
When adding SIC to SINR diagrams, the resulting structures, denoted \emph{SIC-SINR diagrams}, become much more complex to present. However, they clearly reveal the benefits of the cancellation method. An example of this idea is illustrated in Figure \ref{fig:intro}. All three parts of the figure depict a network with two transmitters $s_1, s_2$, and two receivers (or points in the plane) $r_1, r_2$, with the requirement that $r_1$ and $r_2$ need to decode the signal transmitted by $s_1$ and $s_2$ respectively. The four nodes occur on a straight line in the order $\langle r_1, r_2 , s_2, s_1 \rangle$, similar to the known ``nested links" example given in \cite{MWW06}. This example shows that in order to achieve the requirements, a \emph{nonuniform} power assignment must be used by the two transmitters, thus demonstrating that the capacity (achievable rate) of nonuniform power assignments is higher than that of uniform power assignments.
Figure \ref{fig:intro}(a) shows the reception zones $\HH(\Station_1)$ and $\HH(\Station_2)$ for $s_1$ and $s_2$ respectively in the nonuniform setting, which satisfy $r_1 \in \HH(\Station_1)$ and $r_2 \in \HH(\Station_2)$.
As mentioned, it can be proved that the two demands of the system cannot be satisfied
when both $s_1$ and $s_2$ transmit with the same power.
An SINR diagram with a uniform power assignment
is shown in Figure \ref{fig:intro}(b).
Note that here, $r_1 \notin \HH(\Station_1)$, but $r_1 \in \HH(\Station_2)$.
In contrast, when SIC can be employed at $r_1$, it can first decode $s_2$.
Afterwards, it ``cancels" $s_2$ from its received combined signal and then decodes $s_1$. It follows that with SIC the two demands can be satisfied even with uniform powers! The SIC-SINR diagram presented in Figure \ref{fig:intro}(c) illustrates this by showing an additional zone, $\HH(\Station_2,\Station_1)$, the zone in which stations with SIC can decode $s_1$ after ``canceling" $s_2$.
Note that, as explained later, $\HH(\Station_2,\Station_1)$  is the intersection of two convex shapes, $\HH(\Station_2)$ and $\mathcal{H}^*(\Station_1)$, where the latter (shown as an empty circle) is the reception zone of $s_1$ if it had transmitted alone in the network. One clearly sees that the total reception area of $s_1$ with SIC is considerably larger than without SIC.
In Subsection \ref{motivating_example} we present an even more compelling and general
motivating example, that shows the following.
\begin{observation}
\label{obs:motivating_example}
There exists a wireless network for which any power assignment requires
$n$ time slots to satisfy all the demands, while using SIC allows
a satisfying schedule using a \emph{single} time slot.
\end{observation}


Despite the importance of IC, not much is known about its complexity. The goal of this paper is to take a first step towards understanding it, by studying reception maps under the setting of SIC.
The starting point of our work is the observation that under the SIC setting, reception zones are no longer guaranteed to be convex, fat or even connected. This holds even for the ``simplified'' setting where stations transmit at the same power level and are aligned on a straight line (one dimensional map).
The zones are also not hyperbolically convex as was shown for the nonuniform power setting without IC \cite{KLPP2011STOC}.
Moreover, while for SINR diagrams without IC there is a single polynomial that represents each of their reception zones, with IC the reception zone of each transmitter may depend on the cancellation order, which can lead to an exponential number of polynomials and cells. If this were the case, then even drawing the diagram might prove to be infeasible.


\subsection{Main Contributions.}
The study of SIC-SINR reception maps
raises several immediate questions. The first is a simple ``counting'' question that has strong implications on algorithmic issues:
What is the maximum number of reception cells that may occur
in an SINR diagram of a wireless network with $n$ stations, where every point in the map is allowed to perform SIC? Is it indeed exponential?
We address this question in two different ways.
Initially we re-explore the intimate connections between wireless communication and computational geometry methods like higher-order Voronoi diagrams \cite{ShamosH75,ABKS92}.
In particular, we use a bound on the number of cells in \emph{ordered order-$k$ Voronoi diagram} \cite{ABKS92} to upper bound the number of reception zones by
 $O(n^{2d})$, where $n$ is the number of transmitters and $d$ is the dimension.
In general, this bound is not tight, but interestingly we were able to tie the number of reception zones to a novel parameter of the network, termed the {\em Compactness Parameter}, $\CompactnessParameter$, and achieve a much tighter bound when the compactness parameter is sufficiently high.
The compactness parameter is a function of the two most important parameters of the wireless network model, namely, the \emph{reception threshold} constant $\beta \geq 1$, which stands for the SINR threshold, and the {\em path-loss parameter} $\alpha>0$, and its value is $\CompactnessParameter=\sqrt[\alpha]{\beta}$.
We then prove that when $\CompactnessParameter \ge 5$, the number of reception zones is linear for any dimension!
This bound allows us to provide an efficient scheme for computing the cancellation order that gives the reception zones and therefore allows us to build and represent the diagram efficiently.


Our second question has a broader scope: Are there any ``niceness" properties
of reception zones that can be established in the SIC settings?
Specifically, we aim toward finding forms of convexity satisfied by
reception cells in SIC reception maps.
Apart from their theoretical interest, these questions also have considerable practical significance, since having reception zones with some form of convexity might ease the development
of protocols for various design and communication tasks \cite{KLPP2011STOC}.
We answer this question by using the key observation that zones are intersections of convex shapes, giving us some ``nice" geometric guarantees.

Our third question is of algorithmic nature. We consider the \emph{point location}
task, where given a point $p\in \R^{d}$ and a station $\Station_i$, one wants to know whether a receiver lacated at $p$ can receive $\Station_i$'s transmission using SIC. Applying the trivial computation in $O(n \log n)$ time, one can compute the set of stations that $p$ receives under the SIC setting. However, if the number of queries is large, an order of $O(n \log n)$ time per query might be too costly.
To approach this problem we use the guarantees produced for the first two questions and present a scheme for answering point location queries approximately in logarithmic time.

We believe that the questions raised herein, as well as the results and techniques developed, can significantly contribute to the evolving topic of wireless topology and what we may refer to as computational wireless geometry.

The rest of the paper is organized as follows. In Section \ref{section:Uniform_SIC}, we establish the basic properties of SIC-SINR diagrams and show its relation to higher-order Voronoi diagrams.
\commful
We then
\commfulend
\commabs
In Subsection \ref{section:Uniform_Number_Cells}, we
\commabsend
derive a tight bound on the number of connected components in the reception map of a given station under SIC. Section \ref{sec:Cancellation Tree} describes how one can construct SIC-SINR reception maps in polynomial time. Finally, Section \ref{sec:Point_location} considers the point-location task and provides an efficient construction of a data structure that answers point-location queries (with predefined approximation guarantees) in logarithmic time.

\section{Preliminaries}
\label{section:Preliminaries}

\subsection{Geometric notions.}
We consider the $d$-dimensional Euclidean space \(\Reals^d\) (for
$d \in \Integers_{\geq 1}$).
The \emph{distance} between points \(p\) and \(q\) is denoted by
\( \dist{p,q}
= \| q - p \|_2 \).
A \emph{ball} of radius \(r\) centered at point \(p \in \R^{d}\)
is the set of all points
at distance at most \(r\) from \(p\), denoted by
\( \Ball^{d}(p, r) = \{ q \in \Reals^d \mid \dist{p, q} \leq r \} \).
Unless stated otherwise, we assume the 2-dimensional Euclidean plane,
and omit $d$.
The maximal and minimal distances between a point \(p\) and a set of points \(Q\)
are defined, respectively, as \( \maxdist{p,Q}) =\max_{q \in Q} \{\dist{p,q}\} \) and
\( \mindist{p,Q}) =\min_{q \in Q} \{\dist{p,q}\} \).
The hyperplane $HP(q_i,q_j)$, for $q_i,q_j \in \R^{d}$,  is defined by
$HP(q_i,q_j) ~=~ \{p \in \R^d \mid \dist{p,q_i} = \dist{p,q_j}\}$.
Given a set of $n$ points $Q=\{q_{i} \in \R^{d}\}$, let
the corresponding set of all $n \choose 2$ hyperplanes be
$\mathcal{HP}(Q)=\{HP(q_i,q_j) \mid q_i,q_j \in Q\}$.
A finite set $\Upsilon$ of hyperplanes defines a dissection of $\R^{d}$ into
connected pieces of various dimensions, known as the \emph{arrangement}
$\Arrangment(\Upsilon)$ of $\Upsilon$.
The basic notions of open, closed, bounded, compact and connected sets of points are defined in the standard manner (see \cite{Avin2009PODC+full}).

\commabs
{\bf Check if the following detailed definitions are necessary:\\
MP: I put in comment what can be removed } \\
A \emph{maximal connected subset} \(P_1 \subseteq P\) is a
connected point set such that $P_1 \cup \{p\}$ is no longer connected
for every $p \in P \setminus P_1$.
\commabsend

We use the term \emph{zone} to describe a point set with some
``nice'' properties.
Unless stated otherwise, a zone refers to the union of an open connected set
and some subset of its boundary.
It may also refer to a single point or to the finite union of zones.
A polynomial $F: \R^{d} \to \R$ is the \emph{characteristic polynomial}
of a zone \(Z\) if $p\in Z ~\Leftrightarrow~ F(p)\leq 0$ for every $p\in\R^d$.

Denote the \emph{area} of a bounded zone \(Z\)
(assuming it is well-defined) by \(\Area(Z)\).
A nonempty bounded zone \(Z\ne\emptyset\) is \emph{fat}
if the ratio between the radii of the smallest circumscribed and largest inscribed circles
with respect to $Z$ is bounded by a constant.
%
\subsection{Wireless Networks and SINR.}
\label{section:WirelessNetworks}
We consider a wireless network \( \cA = \langle d, S, \Power, \Noise, \beta, \alpha \rangle \), where $d \in \Integers_{\geq 1}$,
\( S = \{ \Station_1, \Station_2, \dots, \Station_{n} \} \) is a set of
transmitting \emph{radio stations} embedded in $d$-dimensional space,
\(\Power\) is a mapping assigning a positive real \emph{transmitting power}
\(\Power_i\) to each station \(\Station_i\),
\( \Noise \geq 0 \) is the \emph{background noise},
\( \beta > 1 \) is a constant serving as the \emph{reception threshold}
(to be explained soon), and $\alpha>0$ is the {\em path-loss parameter}.
The \emph{signal to interference \& noise ratio (SINR)} of \(\Station_i\)
at point \(p\) is defined as
$$\SINR_{\cA}(\Station_{i},p)=\frac{\Power_i \cdot \dist{\Station_i, p}^{-\alpha}}{\sum_{j \neq i}
\Power_j \cdot \dist{\Station_j, p}^{-\alpha} + \Noise}.$$
When the network \(\cA\) is clear from the context, we may omit it and write
simply $\SINR(\Station_{i},p)$.
Throughout this paper we assume the uniform setting, where $\Power=\overline{1}$.
Let \(\cA_{d'}\) be a network identical to \(\cA\) except its dimension
is $d' \neq d$.
In our arguments, we sometimes refer to an ordered subset of stations,
$\orderedSi=(\Station_{i_1}, \ldots, \Station_{i_k}) \subseteq S$.
Denote the last element in $\orderedSi$ by $\LAST(\orderedSi)$.
When the order is insignificant, we refer to this set as simply
$S_i=\{\Station_{i_1}, \ldots, \Station_{i_k}\}$.
The wireless network restricted to a subset of nodes $S_{i}$ is given by
\( \cA(S_{i}) = \langle d, S_{i}, \Power, \Noise, \beta,\alpha \rangle \).
The network is assumed to contain at least two stations, i.e., \( n \geq 2 \).
The fundamental rule of the SINR model is that the transmission of station
\(\Station_i\) is received correctly at point \( p \notin S \) if and only if
its SINR at \(p\) reaches or exceeds the reception threshold of the network,
i.e., $$\SINR(\Station_i, p) \geq \beta.$$
When this happens, we say that \(\Station_i\) is \emph{heard} at \(p\).
\subsection{SINR diagrams (without SIC).}
Let us now introduce the central notion of SINR maps.
We refer to the set of points that hear station \(\Station_i\) as the
\emph{reception zone} of \(\Station_i\), defined as
$$\HH_{\cA}(\Station_i) ~=~
\{ p \in \Reals^d - S \mid \SINR_{\cA}(\Station_i, p) \geq \beta \}
\cup \{\Station_i\} ~.$$
(Note that \(\SINR(\Station_i, \cdot)\)
is undefined at points in \(S\) and in particular at \(\Station_i\) itself.)
Analogously, the set of points that hear none of the stations \(\Station_i \in S\)
(due to the background noise and interference) is defined as
$$ \HH_{\cA}(\emptyset)~=~
\{ p \in \Reals^d - S \mid \SINR(\Station_i, p) < \beta,
~~\forall \Station_i \in S\}.$$
An SINR diagram
$$\ReceptionZone(\cA) = \left(\bigcup_{\Station_i \in S}\HH_{\cA}(\Station_i) \right) \cup \HH_{\cA}(\emptyset)$$ is a ``reception map'' characterizing the reception zones of the stations.
This map partitions the plane into $n+1$ zones; a zone \(\HH_{\cA}(\Station_i)\) for each station $s_i$,
\( 1 \leq i \leq n \), and the zone
\(\HH_{\cA}(\emptyset)\) where none of the stations is received.
It is important to note that a reception zone \(\HH_{\cA}(\Station_i)\)
is not necessarily connected. A \emph{maximal connected component} within
a zone is referred to as a \emph{cell}. Let \(\HH_{\cA}(\Station_i,j)\)
denote the $j^{th}$ cell in \(\HH_{\cA}(\Station_i)\).
\commabs
\par
\commabsend
Hereafter, the set of points where the transmissions of a given station
are successfully received is referred to as its {\em reception zone}.
Hence the reception zone is a set of cells, given by
$$\HH_{\cA}(\Station_i)=\{\HH_{\cA}(\Station_i,1), \ldots
\HH_{\cA}(\Station_i,\NZones_{i})\},$$
where \( \NZones_i = \NZones_i(\cA) \)
is the number of cells in \(\HH_{\cA}(\Station_i)\).
Analogously, \(\HH_{\cA}(\emptyset)\) is composed of
\(\NZones_{\emptyset}(\cA)\) connected cells \(\HH_{\cA}(\emptyset,j)\), for $1 \leq j \leq \NZones_{\emptyset}(\cA)$.
Overall, the topology of a wireless network \(\cA\) is arranged
in three levels:
The \emph{reception map} is at the top of the hierarchy.
It is composed of $n$ reception zones, $\HH_{\cA}(\Station_i)$,
$\Station_i \in S$, and $\HH_{\cA}(\emptyset)$.
Each zone \(\HH_{\cA}(\Station_i)\) is composed of $\NZones_{i}(\cA)$
reception cells.
The following lemma is taken from \cite{Avin2009PODC+full}.
\begin{lemma}[\cite{Avin2009PODC+full}]
\label{fc:uniform_zones}
Let \( \cA = \langle d,
S, \Power, \Noise, \beta, \alpha \rangle \)
be a uniform ($\Power=\overline{1}$) power network where $\beta>1$. Then
\(\HH_{\cA}(\Station_i)\) is convex and fat for every $\Station_i \in S$.
\end{lemma}
In our arguments, we may
sometimes refer to the wireless network $\cA$ induced on a subset
of stations $S_{j} \subseteq S$.
The reception zone of $\Station_i$ in this induced network is denoted by
$\HH_{\cA}(\Station_i ~\mid~ S_{j})$.  When $\cA$ is clear from context,
we may omit it and write simply $\HH(\Station_i)$ and $\HH(\Station_i \mid S_{j})$.
The following definition is useful in our later arguments. Let $F_{\Station_i,\cA}(p)$,
$p\in \R^{d}$, be the \emph{characteristic polynomial} of
$\HH_{\cA}(\Station_i)$, given by
\begin{equation}
\label{eq:reception_polynomial}
F_{\Station_i,\cA}(p) ~=~ \beta \left( \sum_{k \neq i}
\Power_{k} \prod_{l \neq k} \dist{\Station_{l},p}^{\alpha} +
\Noise\prod_{k} \dist{\Station_{k} ,p}^{\alpha}\right)
- \Power_{i}\prod_{k \neq i} \dist{\Station_{k},p}^{\alpha}~.
\end{equation}
Then $p \in \ReceptionZone_{i}(\cA)$ if and only if $F_{\Station_i,\cA}(p) \leq 0$.

\subsection{Geometric diagrams in $\R^d$.}
Throughout the paper we make use of the following types of diagrams.
\subsubsection*{Hyperplane Arrangements.}
Given a set of $\Upsilon$ of $n$ hyperplanes in $\R^{d}$, the arrangement $\Arrangment(\Upsilon)$
of $\Upsilon$ dissects $\R^{d}$ into connected pieces of various dimensions.
Let $\NVorZonesArrangment(\Upsilon)$ denote the number of connected components
in $\Arrangment(\Upsilon)$. The following facts about $\Arrangment(\Upsilon)$
are taken from \cite{Edelsbrunner-CG}.
\begin{lemma}[\cite{Edelsbrunner-CG}]
\label{lem:arrangments}
(a)  $\NVorZonesArrangment(\Upsilon)=\Theta(n^{d})$.\\
(b) $\Arrangment(\Upsilon)$ can be constructed in $\Theta(n^{d})$ time
and maintained in $\Theta(n^{d})$ space.
\end{lemma}
Given a set of $n$ points $S \subset \R^{d}$, we define $\Arrangment(S)$ to be
the arrangement on $\mathcal{HP}(S)=\{HP(s_{i},s_{j}) \mid s_i,s_j \in S\}$,
the set of all $n \choose 2$ hyperplanes of pairs in $S$.
$\Arrangment(S)$ has an important role in constructing SIC-SINR  maps,
as will be described later on.
\begin{corollary}
\label{cor:Sarrangments}
$\NVorZonesArrangment(S)=\Theta(n^{2d})$.
\end{corollary}

\subsubsection*{Voronoi diagrams.}
The ordinary Voronoi diagram on a given set of points $S$ is generated by assigning each point in the space to the closest
point in $S$, thus partitioning the space into {\em cells},
each consisting of the set of locations closest to one point in $S$
(referred to as the cell's {\em generator}).
Let $\vor(\Station_i)$ denote the Voronoi cell of $\Station_i$ given a set
of generators $S$.
Let $\vor(\Station_i \mid S_{j})$, for $S_{j} \subseteq S$, denote the
Voronoi cell of $\Station_i$ in a system restricted to the points of $S_{j}$.


Avin et al. \cite{Avin2009PODC+full} discuss the relationships between
the SINR diagram on a set of stations $S$ with \emph{uniform} powers and the
corresponding {\em Voronoi diagram} on $S$, and establishes the following lemma.
Let \( \cA= \langle d, S , \overline{1}, \Noise, \beta \geq 1,\alpha \rangle\).
\begin{lemma}[\cite{Avin2009PODC+full}]
\label{lemma:Voronoi_SINR_WOSIC}
$\HH_{\cA}(\Station_i) \subseteq \vor(\Station_i)$
for every $\Station_i \in S$.
\end{lemma}
\subsubsection*{Higher order Voronoi diagrams.}
Higher order Voronoi diagrams are a natural extension of the ordinary Voronoi
diagram, where cells are generated by more than one point.
They provide tessellations where each region consists of the locations having
the same $k$ (ordered or unordered) closest points in $S$,
for some given integer $k$.
\subsubsection*{Order-$k$ Voronoi diagram.}
The order-$k$ Voronoi diagram $\mathbb V^{(k)}(S)$  is the set of all non-empty order-$k$ Voronoi regions $\mathbb V^{(k)}(S)=\{\vor(S_{1}^{(k)}), \ldots,\vor(S_{m}^{(k)}) \}$, where the order-$k$ Voronoi zone $\vor(S_{i}^{(k)})$ for an unordered subset $S_{i}^{(k)} \subseteq S, |S_{i}^{(k)}|=k$, is defined as follows.
\begin{eqnarray*}
\vor(S_{i}^{(k)}) = \{p \in \R^{d}  &\mid& \maxdist{p,S_{i}^{(k)}})\leq \mindist{p, S \setminus S_{i}^{(k)}})\}~.
\end{eqnarray*}
This can alternatively be written as
\begin{equation}
\label{eq:UnorderedVoronoiChain}
\vor(S_{i}^{(k)})=\bigcap_{s \in S_{i}^{(k)}} \vor(s \mid S \setminus S_{i}^{(k)} \cup \{s\})~.
\end{equation}
This alternate representation plays a role in this paper.
Note that $\mathbb V^{(1)}(S)$ corresponds to the ordinary Voronoi diagram
and that any $\mathbb V^{(k)}(S)$, for $k>1$, is a refinement of $\mathbb V^{(1)}(S)$.
\subsubsection*{Ordered Order-$k$ Voronoi diagram.}
Let $\orderedSi\subseteq S$ be an ordered set of $k$ elements from $S$.
When the $k$ generators are ordered, the diagram becomes the
\emph{ordered order-$k$ Voronoi diagram}
$\mathcal{V}^{ \langle k \rangle}(S)$ \cite{ABKS92}, defined as
$$\mathcal{V}^{\langle k \rangle}(S)=\{\vororder{S_{i}}\},$$
where the ordered order-$k$ Voronoi region $\vororder{S_{i}}$, $|\orderedSi|=k$,
is defined as
\begin{eqnarray*}
\vororder{S_{i}} &=& \{p \in \R^{d} \mid \dist{p,\Station_{i_{1}}}
\leq \dist{p,\Station_{i_{2}}} \leq \ldots
\leq \dist{p,\Station_{i_{k}}}
\leq \mindist{p,S \setminus S_{i}})\}.
\end{eqnarray*}
Alternatively,
\commabs
it can written
\commabsend
as in \cite{ABKS92},
\begin{equation}
\label{eq:OrderedVoronoiChain}
\vororder{S_{i}} ~=~ \bigcap_{j=1}^{k} \vor(\Station_{i_{j}} \mid
S \setminus \{\Station_{i_{1}}, \ldots, \Station_{i_{j-1}}\})~.
\end{equation}
Note that each $\vororder{S_{i}}$ is an intersection of $k$ convex shapes and hence it is convex as well.\\
The following claim is useful for our later arguments.
\begin{lemma}
\label{lem:non_intersected_cells}
For every $\orderedSi,\orderedSj \subseteq S$ such that
$\orderedSi \nsubseteq \orderedSj$ there exist
$\Station_{k_1},\Station_{k_2} \in S$, such that the hyperplane
$HP(\Station_{k_1},\Station_{k_2})$ separates $\vororder{S_{i}}$
and $\vororder{S_{j}}$.
\end{lemma}
\Proof
We focus on the case where $\vororder{S_{i}} \neq \emptyset$ and
$\vororder{S_{j}} \neq \emptyset$. Let $m$ denote the first index such that
$\Station_{i_{m}} \neq \Station_{j_{m}}$.
First consider the case where $m=1$.
Then $\vororder{S_{i}} \subseteq \vor(\Station_{i_{1}})$
and $\vororder{S_{j}}\subseteq \vor(\Station_{j_{1}})$,
so $HP(\Station_{i_{1}},\Station_{j_{1}})$ separates the zones and the lemma holds.
Otherwise, assume $m>1$ and let
$\overrightarrow{S^*}=\{\Station^{*}_{1}, \ldots, \Station^{*}_{m-1}\}$
denote the longest common prefix of $\orderedSi$ and $\orderedSj$.
Let $X_0=\vororder{S^{*}}$,
$X_1=\vor(\Station_{i_{m}} \mid S \setminus S^{*})$ and
$X_2=\vor(\Station_{j_{m}} \mid S \setminus S^{*})$.
First note that by Eq. (\ref{eq:OrderedVoronoiChain}),
$\vororder{S_{i}} \subseteq X_0 \cap X_1$ and
$\vororder{S_{j}} \subseteq X_0 \cap X_2$.
In addition, $X_0, X_1$ and $X_{2}$ are convex.
Next, observe that $X_1, X_2$ correspond to distinct Voronoi regions in the
system of points $S \setminus S^{*}$ and therefore $X_1$ and $X_2$
are separated by $HP(\Station_{i_{m}},\Station_{j_{m}})$. The lemma follows.
\QED

\subsection{Motivating Example: Interference Cancellation vs. Power Control.}
\label{motivating_example}
\begin{figure}[h!]
\begin{center}
\includegraphics[scale=0.45]{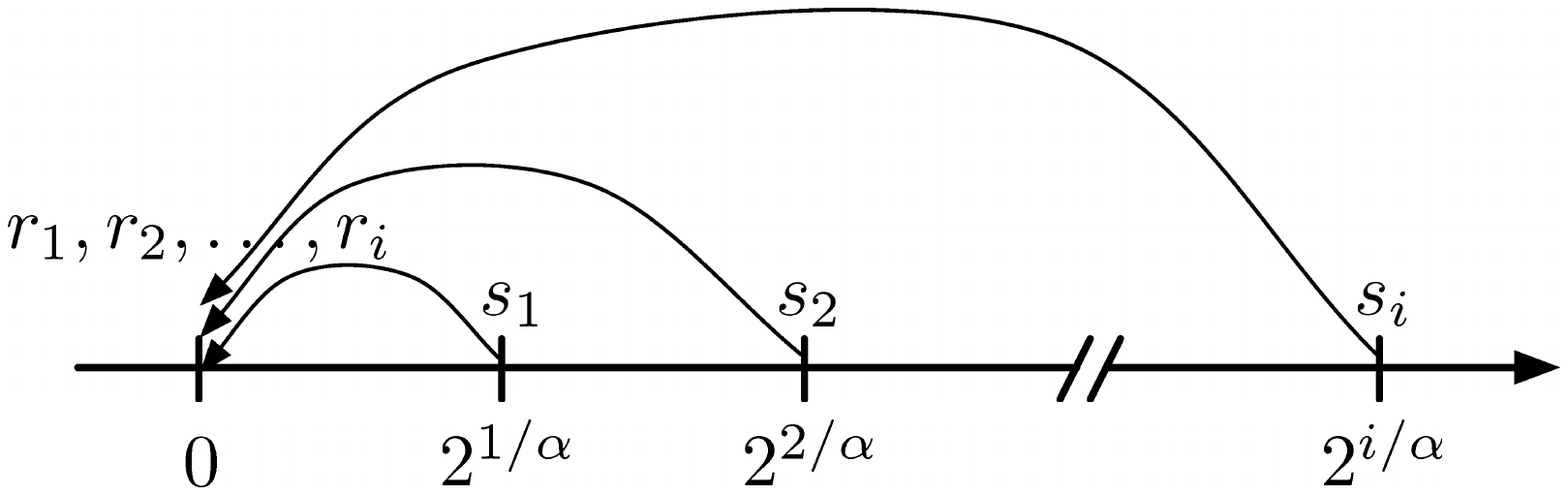}
\caption{ \label{figure:exponential_chain}
\sf The power of interference cancellation.
Any power assignment requires $n$ time slots to satisfactorily schedule these $n$ requests in the SIC-free setting. Using SIC, all requests are satisfied in a single time slot (even when using \emph{uniform} powers).
}
\end{center}
\end{figure}

\commful
The following motivating example
\commfulend
\commabs
We initiate our study by providing a motivating example that
\commabsend
illustrates the power of interference cancellation even in the uniform setting
where all stations use the same transmission power.
Consider a set of $n$ communication requests $L=\left \{\left(\Station_{i},r_{i}\right)\mid i=1,...,n \right\}$ consisting of sender-receiver pairs embedded on a real line as follows. The $n$ receivers are located at the origin and the set of senders are positioned on an exponential node-chain, e.g., $\Station_i$ is positioned on $x_{i}=2^{i/\alpha}$ (see Fig \ref{figure:exponential_chain}).  Since all receivers share the same position, without SIC there exists no power assignment that can satisfy more than one request simultaneously, hence $n$ time slots are necessary for satisfying all the requests.
We claim that by using SIC, all requests can be satisfied in a single time slot even with a uniform power assignment. We focus on a given receiver $r_{j}$ and show that it successfully decodes the signal from $s_{j}$ after successive cancellations of the signal transmitted by $s_i$ for every $i <j$.
Using the notation of Section \ref{section:WirelessNetworks}, let
\( \cA_{i} = \langle d, S_{i}=\{\Station_{i}, \ldots,\Station_{n}\},
\overline{1}, \Noise\leq  1/2^{n}, \beta=1,\alpha \rangle \)
denote the network imposed on the last $n-i$ stations,
whose positions are $2^{(n-i)/\alpha}$ to $2^{n/\alpha}$.
Note that
$$\SINR_{\cA_{i}}(\Station_i,r_{j}) ~=~
\frac{1/2^{i}}{\sum_{j=i+1}^{n} 1/2^{j}+\Noise} \geq 1$$
for every $i \leq n$.
We therefore establish that there exists an instance
$L=\left\{\left(s_i,r_i \right) \mid i=1,...,n \right\}$
such that any power assignment for scheduling $L$ requires $n$ slots, whereas using SIC allows a satisfying schedule using a \emph{single} time slot.

So far, the literature on capacity and scheduling addressed mostly nonuniform
powers, showing that nonuniform power assignments can outperform a
uniform assignment \cite{MWW06,MoWa06} and increase the capacity of a network.
In contrast, examples such as Fig. \ref{figure:exponential_chain} illustrate the power of interference cancellation even with uniform power assignments, and motivate the study of this technique from an algorithmic point of view.
Understanding SINR diagrams with SIC may play a role in the development of suitable algorithms (e.g. capacity, scheduling and power control), filling the current gap between the electrical engineering and algorithmic communities with respect to SIC research.

\section{SIC-SINR Diagrams in Uniform Power Networks}
\label{section:Uniform_SIC}
In this section, we first formally define the reception zones under
interference cancellation, forming the \emph{SIC-SINR Diagrams}.
We then take a first step towards studying the properties of these diagrams.
We elaborate on the relation between the SIC-SINR diagram and the
\emph{ordered order-k Voronoi diagram}, and use it to prove convexity
properties of the diagram and to bound the number of connected components
in the SIC-SINR Diagrams. We then define the \emph{Compactness Parameter}
of the diagrams and use it to achieve tighter bounds
on the number of connected components.
\subsection{SINR diagrams with SIC.}\label{subsec:ICnot}
Let \( \cA = \langle d, S, \Power=\overline{1}, \Noise, \beta > 1, \alpha \rangle \).
We now focus on the reception zone of a single station, say $\Station_1$,
under the setting of interference cancellation.
In other words, we are interested in the area containing all points that can decode $\Station_1$, possibly after some sequence of successive cancellations.
To warm up, we start with the case of a single point $p\in \R^{d}$ and ask
the following question: does $p$ successfully receive $\Station_1$ using SIC?

Let $\orderedSp=\{\Station_{p_1}, \ldots, \Station_{p_k}\}$ correspond to
the set of stations $S$  ordered in nonincreasing order of received
signal strength at point $p$ up to station $s_1$, i.e.,
$\Energy_{\cA}(\Station_{p_1}, p) \geq \Energy_{\cA}(\Station_{p_2}, p)\geq
\ldots \geq \Energy_{\cA}(\Station_{p_k}, p)$, where
$\Station_{p_k}=\Station_1$.
Since all stations transmit with the same power, it also holds that
\begin{equation}
\label{eq:dist-order}
\dist{\Station_{p_1}, p} ~\leq~ \dist{\Station_{p_2}, p} ~\leq~ \ldots
~\leq~  \dist{\Station_{p_k}=s_1, p}.
\end{equation}
To receive $\Station_1$ correctly, $p$ must successively cancel the signals transmitted by station $\Station_{p_i}$, for $i <k$, from the strongest signal to the weakest. It therefore follows that $p$ successfully receives $\Station_1$ following SIC iff
\begin{equation}
\label{eq:SIC_condition}
p \in \HH(\Station_{p_i} \mid S \setminus
\{\Station_{p_1}, \ldots,\Station_{p_{(i-1)}}\}),
\end{equation}
for every $i\leq k$.
The reception zone of $\Station_1$ in a wireless network \(\cA\)
under the setting of SIC is denoted by $\HH^{SIC}_{\cA}(\Station_{1})$,
or simply $\HH^{SIC}(\Station_{1})$ when $\cA$ is clear from the context.
It contains $\Station_1$ and the set of points $p$ obeying Equation
(\ref{eq:SIC_condition}), i.e.,
\begin{equation}
\label{eq:SIC_condition_zone}
\HH^{SIC}(\Station_{1}) ~=~ \{p\in \Reals^d - S \mid
\text{$p$ satisfies Eq.} ~(\ref{eq:SIC_condition})\}.
\end{equation}
We now provide a more constructive formulation for $\HH^{SIC}(\Station_{1})$,
which becomes useful in our later arguments. Let $\orderedSi \subseteq S$
be an ordering of $k$ stations $\Station_{i_1}, \ldots, \Station_{i_{k}}$. Let $\HORDER{S_{i}}$ denote
the reception area of all points that receive $\LAST(\orderedSi)$ correctly
after successive cancellation of $\Station_{i_1}, \ldots, \Station_{i_{k-1}}$.
Formally, the zone $\HORDER{S_{i}}$ is defined in an inductive manner
with respect to the length of the ordering $\orderedSi$, i.e.,
number of cancellations minus one. For $\orderedSi=\{\Station_j\}$,
$\HORDER{S_{i}}=\HH(\Station_j)$. Otherwise, for $k >1$,
\begin{equation*}
\HORDER{S_{i}} ~=~ \HH(\orderedSi \setminus \{\Station_{i_{k}}\}) \cap
\HCON(\Station_{i_{k}} \mid (S \setminus S_{i}) \cup \{\Station_{i_{k}}\})~,
\end{equation*}
or
\begin{equation}
\label{eq:unique_order_zone}
\HORDER{S_{i}}=\bigcap_{j=1}^{k} \HCON(\Station_{i_{j}} \mid
S \setminus \{\Station_{i_1}, \ldots,\Station_{i_{(j-1)}}\})~.
\end{equation}
The following is a direct consequence of Eq. (\ref{eq:unique_order_zone}).
\begin{corollary}
\label{cor:SIC_cell}
Let $\orderedSi \subseteq S$, $|\orderedSi|=k$. Then
\begin{eqnarray*}
\HORDER{S_{i}} &\subseteq& \HH(\Station_{i_1}, \ldots, \Station_{i_{(k-1)}})
\subseteq \HH(\Station_{i_1}, \ldots, \Station_{i_{(k-2)}})
\subseteq \ldots \subseteq \HH(\Station_{i_1})~.
\end{eqnarray*}
\end{corollary}
Finally, the reception zone of $\Station_1$ under SIC is given as follows. Let $\CO_j$ denote the collection of all cancellation orderings
ending with $s_j$, namely,
$$\CO_j = \{ \orderedSi \subseteq S \mid \LAST(\orderedSi)=\Station_j \}.$$ Then
\begin{equation}
\label{eq:zone_cancellation}
\HH^{SIC}(\Station_1) ~=~ \bigcup_{\orderedSi\in\CO_1} {\HORDER{S_{i}}}~.
\end{equation}
The reception zone
$$\HH^{SIC}(\Station_1)=\{\HH^{SIC}(\Station_1,1), \ldots
\HH^{SIC}(\Station_1,\NZones_{1}^{SIC})\}$$
is a set of \(\NZones_{1}^{SIC}\) cells.
Although by definition it seems that $\HH^{SIC}(\Station_1)$ might consist of an exponential number of regions $\HORDER{S_{i}}$ for each $\orderedSi\in\CO_1$, in what follows we show that this is not the case and that there are only polynomially many cancellation ordering $\orderedSi\in\CO_1$ that are relevant for $\HH^{SIC}(\Station_1)$.
Note that the region of unsuccessful reception to any of the points, namely,
\(\HH(\emptyset)\), is unaffected by SIC. This follows by noting that SIC
only affects the set of points
$p \in \bigcup \HH(\Station_i)\setminus S$.
In other words, the successive signal cancellation allows points
$p \in \R^{d} \setminus S$ to ``migrate'' from  the reception zone
of station $\Station_i$ to that of station $\Station_j$.
However, points that hear nobody can cancel none of the signals.
Overall, analogous to the SIC-free case, the topology of a wireless network \(\cA\) under SIC is again arranged
in three levels:
The \emph{reception map}, at the top of the hierarchy,
is composed at the next level of $n$ reception zones, $\HH^{SIC}(\Station_i)$,
$\Station_i \in S$ and \(\HH(\emptyset)\).
Finally, at the lowest level, each zone \(\HH^{SIC}(\Station_i)\) is composed of $\NZones_{i}^{SIC}$
reception cells.

\par Analogous to the SIC-free setting, we may refer to the wireless network $\cA$ induced on a subset of stations $S_{j} \subseteq S$. The reception zone $\HORDER{S_{i}}$ in this induced network is denoted by $\HH(\overrightarrow{S}_{i} \mid S_j)$ or  $\HH(\Station_{i_1}, \ldots, \Station_{i_{k}} \mid S_j)$.

\par Throughout the paper we consider a uniform power network of the form
\( \cA = \langle d , S, \Power=\overline{1}, \Noise, \beta > 1,
\alpha \rangle \).
By Lemma \ref{fc:uniform_zones}, reception zones of uniform SIC-free power maps are convex. However, once signal cancellation is allowed,
the convexity (and connectivity) of the zones is lost, even for the simple case
where stations are aligned on a line; see Figure \ref{figure:SIC_Zones_1d}
for an illustration of the SIC-SINR map of a 3-station system.

%
\subsection{Higher-order Voronoi diagrams and SIC-SINR maps.}
\label{section:Vor_SIC}
To understand the structure and the topological properties of SIC-SINR
reception maps, we begin our study by describing the relation between SIC-SINR
reception maps and ordered order-$k$ Voronoi diagram. Specifically, we prove that every SIC-SINR zone is composed of a collection of convex cells, each of which is related to a cell of the higher-order Voronoi diagram.
To avoid complications, we assume our stations are embedded in
\emph{general positions}.
\par We begin by describing the relation between a nonempty reception region
$\HORDER{S_{i}}$ and an nonempty ordered order-$k$ polygon.
\begin{lemma}
\label{lem:Voronoi_SINR_SIC}
For every $\beta \geq 1$, $\HORDER{S_{i}} \subseteq \vororder{S_{i}}$.
\end{lemma}
\Proof
By Lemma \ref{lemma:Voronoi_SINR_WOSIC},
$\HH(\Station_{j} \mid S') \subseteq \vor(\Station_j \mid S')$.
Therefore by Eq. (\ref{eq:unique_order_zone}) it follows that
\begin{eqnarray*}
\HORDER{S_{i}} &\subseteq& \bigcap_{i=1}^{k}  \vor(\Station_{i_{j}} \mid
(S \setminus \{\Station_{i_{1}}, \ldots, \Station_{i_{j-1}}\})= \vororder{S_{i}}~,
\end{eqnarray*}
where the last inequality follows by Eq. (\ref{eq:OrderedVoronoiChain}).
\QED

We now show that reception regions in $\HH^{SIC}(\Station_1)$
that result from different cancellation orderings correspond to
distinct connected cells.
\begin{lemma}
\label{lem:H_S_cells}
Every two regions $\HORDER{S_{1}}, \HORDER{S_{2}} \subseteq
\HH^{SIC}(\Station_1)$ correspond to two distinct cells.
\end{lemma}
\Proof
By Eq. (\ref{eq:zone_cancellation}), $\HH^{SIC}(\Station_1)$ is the union of
$\HORDER{S_{i}}$ regions for $\orderedSi\in\CO_1$,
i.e., where $\LAST(\orderedSi)=\Station_1$.
By Lemma \ref{lem:Voronoi_SINR_SIC}, $\HORDER{S_1} \subseteq \vororder{S_1}$
and $\HORDER{S_2} \subseteq \vororder{S_2}$.
Due to Claim \ref{lem:non_intersected_cells},
$\vororder{S_{1}}\cap\vororder{S_{2}}=\emptyset$ and hence also
$\HORDER{S_1} \cap \HORDER{S_2}=\emptyset$. The lemma follows.
\QED
This lemma establishes the following.
\begin{lemma}
\label{cor:H_S_cells_vor}
For every two reception cells $\HH^{SIC}(\Station_1,i)$ and
$\HH^{SIC}(\Station_1,j)$,
there are distinct orderings
$\orderedSi,\orderedSj  \in \CO_{1}$ such that
$\HH^{SIC}(\Station_1,i) \subseteq \vororder{S_{i}}$ and
$\HH^{SIC}(\Station_1,j) \subseteq \vororder{S_{j}}$.
\end{lemma}
For illustration of these relations, see Figures \ref{figure:SIC_Zones_1d}
and \ref{figure:SIC_Zones_2d}.

\def\FIGC{
\begin{figure*}[h!]
\begin{center}
\includegraphics[scale=0.5]{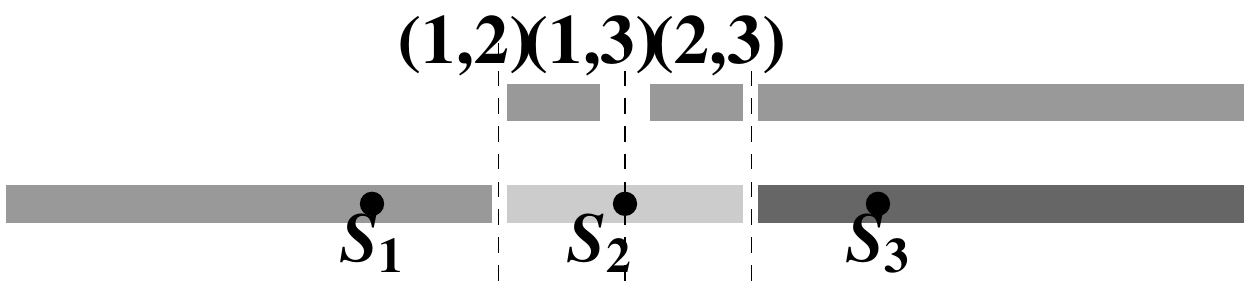}
\hfill
\includegraphics[scale=0.5]{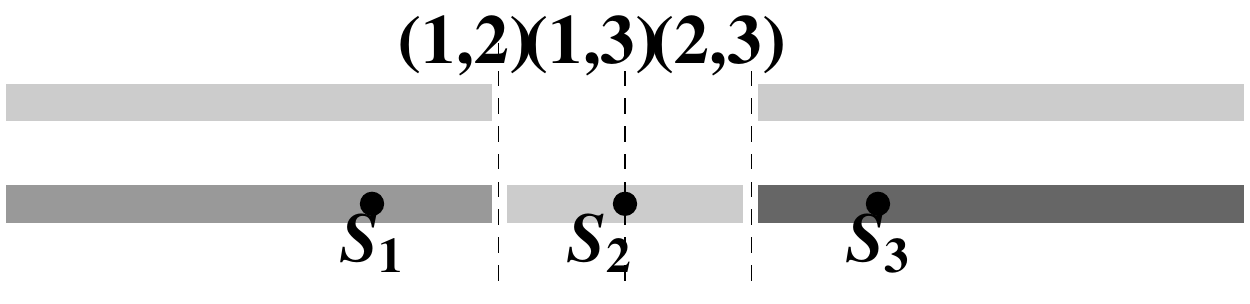}
\hfill
\includegraphics[scale=0.5]{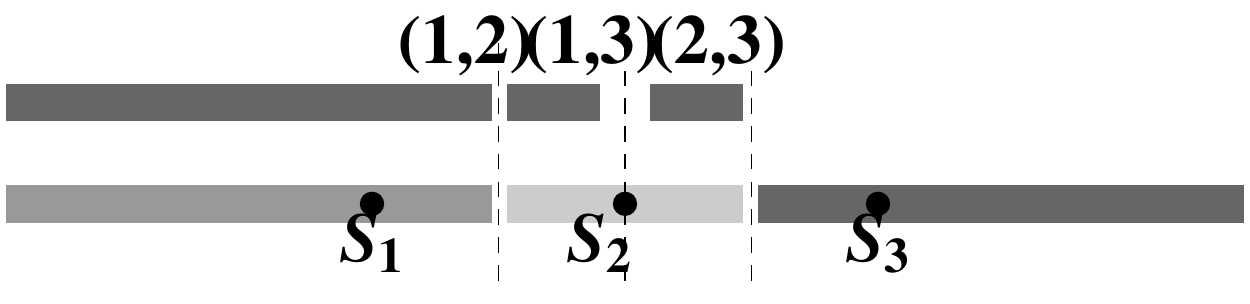}
\caption{ \label{figure:SIC_Zones_1d}
\sf SIC-SINR reception map in $\R^1$ for a 3-station network aligned on a line.
The first line of colored segments corresponds to reception zones with no
cancellations.  The second line of segments in each figure represents the added
reception cells by signal cancellation.
$\HH^{SIC}_{\cA_{d=1}}(\Station_1)$, $\HH^{SIC}_{\cA_{d=1}}(\Station_2)$ and
$\HH^{SIC}_{\cA_{d=1}}(\Station_3)$ are in middle, light and dark grey respectively.
}
\end{center}
\end{figure*}
} 
\FIGC
\def\FIGD{
\begin{figure*}[h!]
\begin{center}
\includegraphics[scale=0.45]{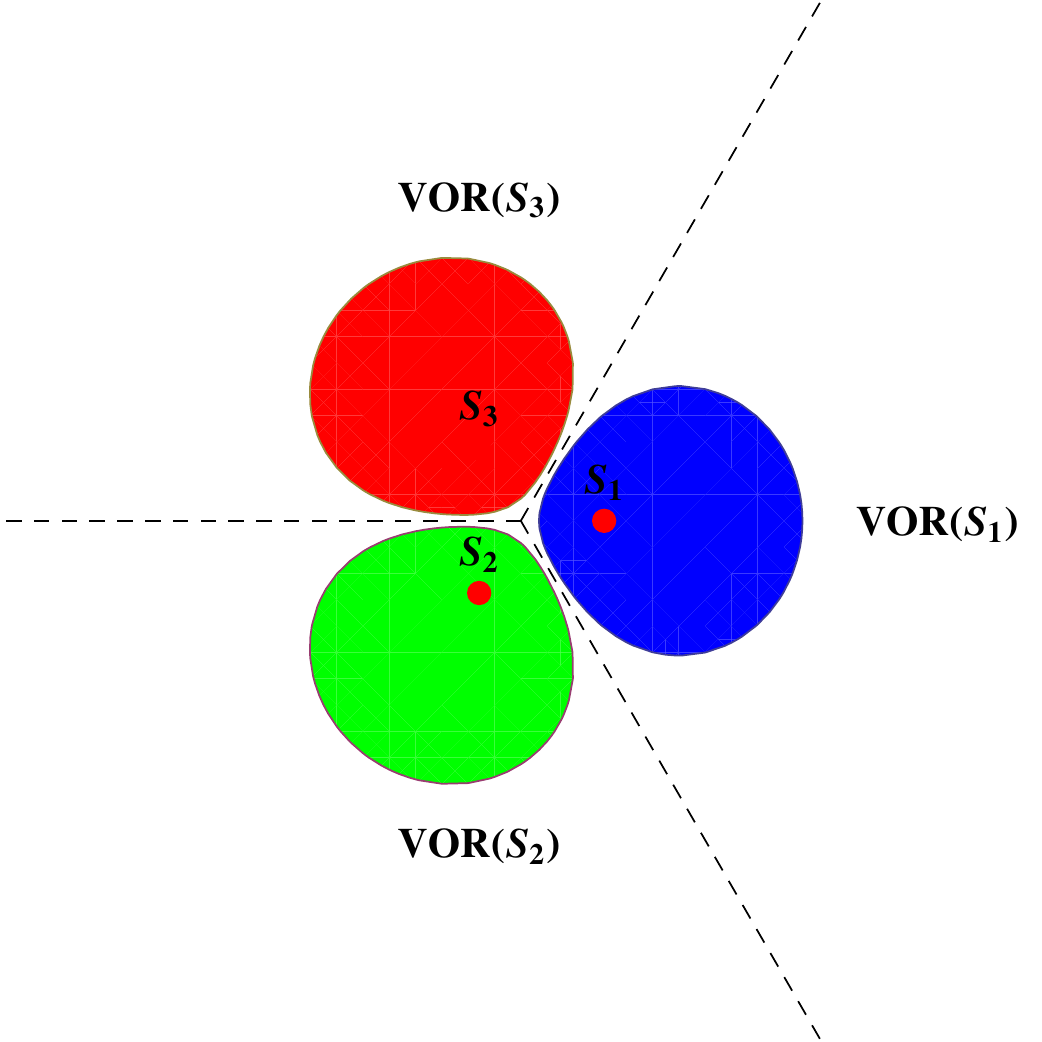}
\hfill
\includegraphics[scale=0.45]{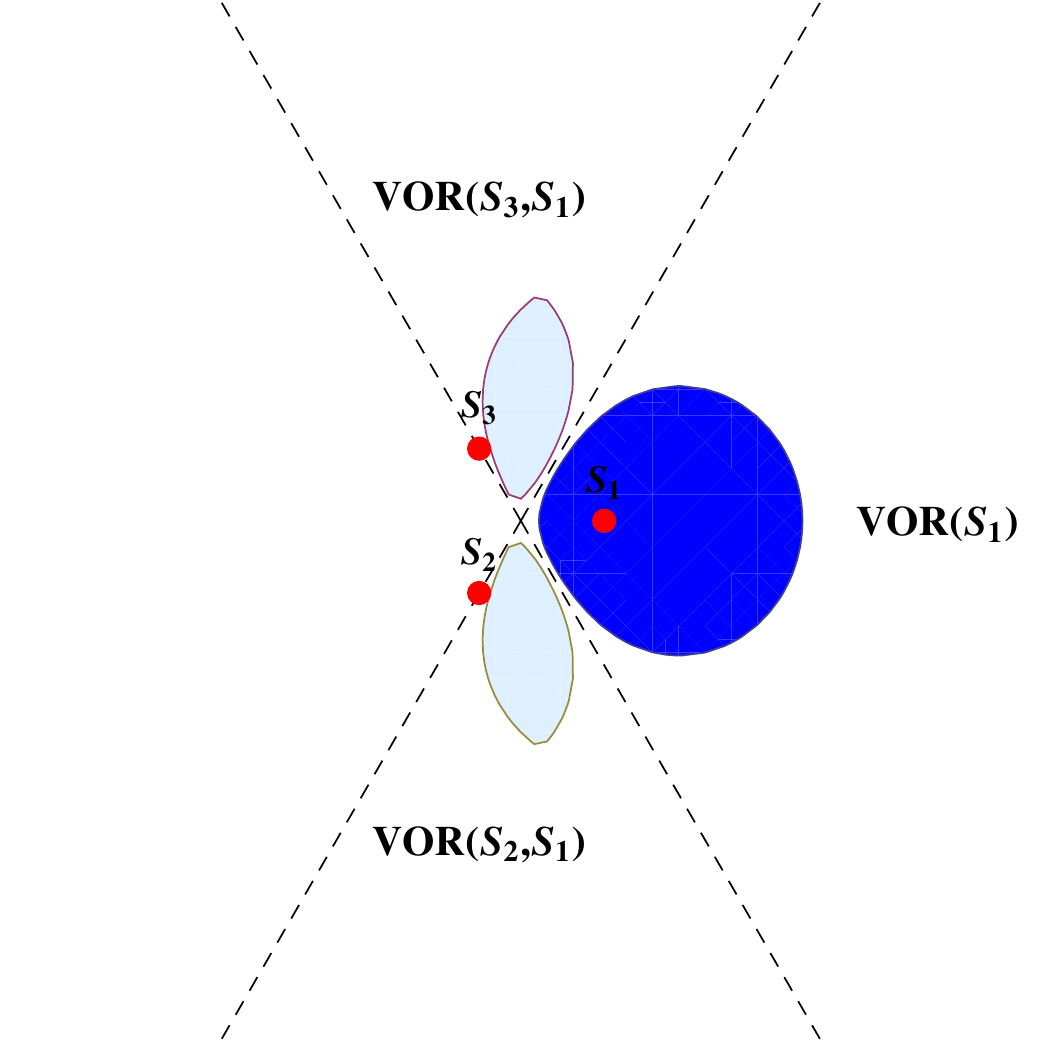}
\hfill
\includegraphics[scale=0.45]{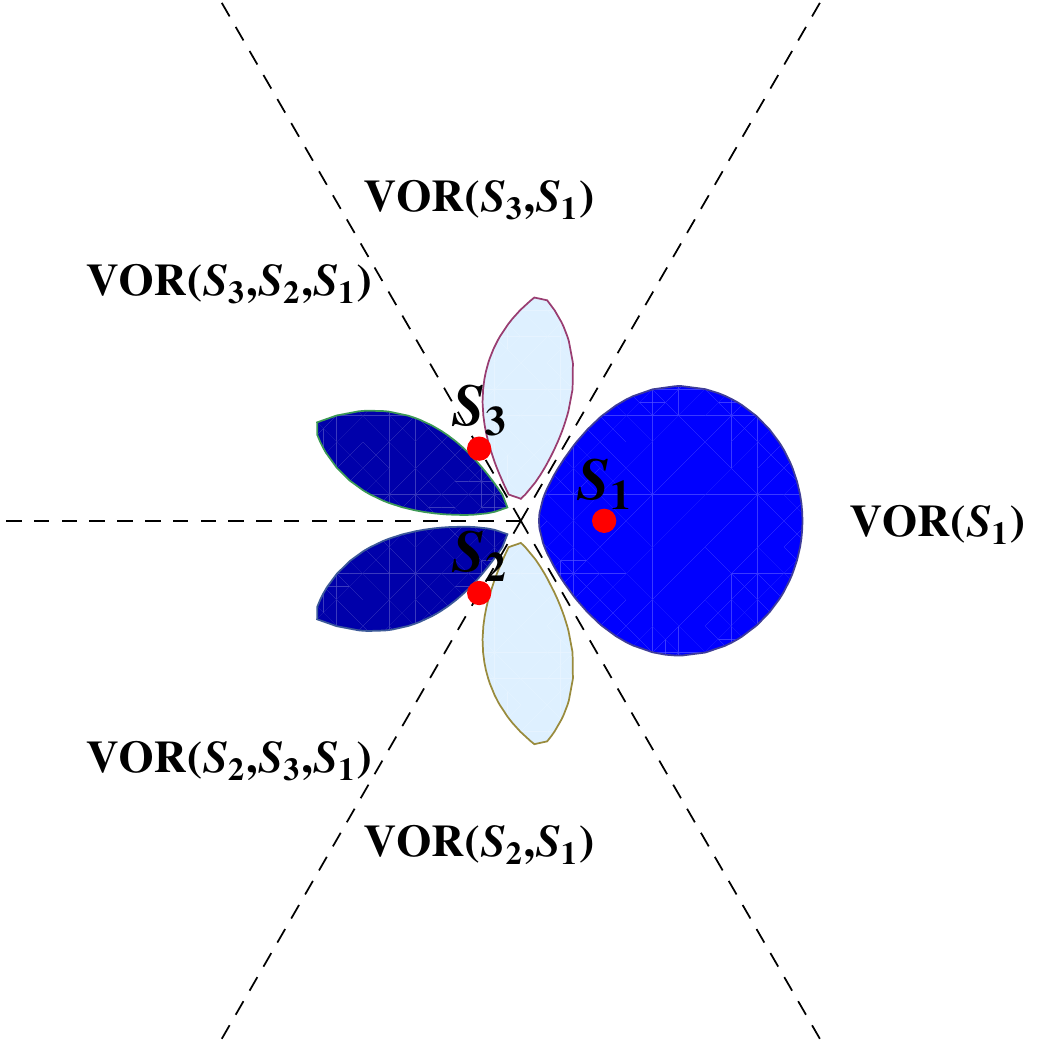}
\caption{ \label{figure:SIC_Zones_2d}
\sf Reception map of $\HH^{SIC}_{\cA_{d=2}}(\Station_1)$ and ordered order-$k$
Voronoi diagram (for $k \in [1,3])$.
(a) SINR map with no cancellation. Shown are $\HH(\Station_1)$,
$\HH(\Station_2)$ and $\HH(\Station_3)$.
(b) Intermediate map of $\HH^{SIC}_{\cA_{d=2}}(\Station_1)$: Reception cells
of $\Station_1$ following at most one cancellation.
(c) Final map of $\HH^{SIC}_{\cA_{d=2}}(\Station_1)$: Reception cells
of $\Station_1$ following two cancellations.
}
\end{center}
\end{figure*}
} 
\FIGD

Next, this relation between the SIC-SINR reception maps and
ordered order-$k$ Voronoi diagram is used to establish the convexity of cells
and to bound the number of connected components in the zone.
We first show that the reception cells of $\HH^{SIC}(\Station_1)$ are convex.
\begin{lemma}
\label{lem:uniform_cells_convex}
Every reception cell $\HH^{SIC}(\Station_1,i), i \in [1, \NZones_{i}^{SIC}],$ is convex.
\end{lemma}
\Proof
Due to Lemma \ref{lem:H_S_cells} it is enough to show that every nonempty $\HORDER{S_{i}}$ is convex. Let $k=|S_i|$ and let $X_j=\HCON(\Station_{i_{j}} \mid S \setminus
\{\Station_{i_1}, \ldots,\Station_{i_{(j-1)}}\})$.
By Lemma \ref{fc:uniform_zones}, $X_j$ is convex for every $j \leq k$
and therefore by Eq. (\ref{eq:unique_order_zone}), $\HORDER{S_{i}}$ is
an intersection of $k$ convex and bounded shapes,
hence it is convex (and bounded) as well.
\QED

We now discuss the number of connected components in SIC-SINR
diagrams. Without loss of generality we focus on station $\Station_1$.
By Lemma \ref{lem:H_S_cells}, every two distinct orderings correspond to distinct reception cells (though it might be empty).
Since the number of distinct orderings of length $n-1$  is $(n-1)!$  (i.e., the size of $\CO_{1}$) and each of those orderings might correspond to a distinct cell, it follows that
the number of connected cells in $\HH^{SIC}(\Station_1)$ might be exponential.
Fortunately, the situation is much better due to Lemma \ref{cor:H_S_cells_vor}.
An ordering $\orderedSi$ is defined as a \emph{nonempty cancellation ordering
($\NCO$)} if and only if $\vororder{S_{i}}$ is nonempty.
Partition the collection of $\NCO$'s into sets $\NCO_{1}, \ldots, \NCO_{n}$
as follows.
An ordering $\orderedSi$ is in the set $\NCO_{j}$  if and only if it is an $\NCO$ and in addition $\LAST(\orderedSi)=\Station_j$.
\par We first consider the hyperplane arrangement
$\Arrangment(S)$ and claim that any given cell $f \in \Arrangment(S)$ intersects with at most one high-order Voronoi region $\vororder{S_{i}}$, where $\overrightarrow{S}_{i}\in \NCO_1$.
\begin{lemma}
\label{cl:high_order_arrangment}
Let $f \in \Arrangment(S)$. Then there exists at most one $\overrightarrow{S}_{i} \in \NCO_1$ such that $f \cap \vororder{S_i} \neq \emptyset$.
\end{lemma}
\Proof
Assume to the contrary that there are two distinct Voronoi regions
corresponding to two orderings, $C_{1}=\vororder{S_{i_1}}$ and
$C_{2}=\vororder{S_{i_2}}$, where $S_{i_1}, S_{i_2} \in \NCO_{1}$,
that have a nonempty intersection with a common face $f$, i.e.,
$C_{1} \cap f \neq \emptyset$ and $C_{2} \cap f \neq \emptyset$.
By Claim \ref{lem:non_intersected_cells}, there exists a hyperplane
$HP(\Station_{q_1},\Station_{q_2}) \in \mathcal{HP}(S)$ that separates $C_1$ and $C_2$.
As $C_1$ and $C_2$ are convex, it follows that this separating hyperplane must
intersect $f$ as well. However, by the definition of arrangements, $f$ is not
intersected by any separating hyperplane, contradiction.
\QED
The following lemma
shows that there are only polynomially many orderings in $\NCO_{j}$.
\begin{lemma}
\label{lem:order_ncells}
(a)~$|\NCO_{1}|=O(n^{2d})$, and (b)~$|\NCO_{1}|=\Omega(n^{2})$ for $d=1$.
\end{lemma}
\Proof
Part (a) follows by combining Lemma \ref{cl:high_order_arrangment} with Corollary \ref{cor:Sarrangments}. To prove Part (b) we provide a construction
of an $n$-station wireless network in $\R^{1}$ that has $\Omega(n^{2})$ cells
corresponding to $\NCO_1$, asymptotically matching the upper bound for $d=1$.
Consider a set of $n$ points $S$ on the line where $s_{i}$ is positioned on $x_{i}$. Select the points $x_1, \ldots, x_{n}$ so that $x_2>x_1$ and $x_{i+1} > x_{i}+x_{i-1}-x_{1}$ (for every $i\geq 2$). We now show that $|\NCO_{1}|=\Omega(n^{2})$.
Let $m_{i,j}$ be the mid point between $s_{i}$ and $s_{j}$, i.e., $m_{i,j} = (x_{i}+x_{j})/2$.
Consider an iterative network construction process in which we start with an empty set of stations $S=\emptyset$, and at step $i$ we add a station $s_{i}$ at location $x_{i}$.
We claim that for each step $i \in \{1, \ldots, n-1\}$  the following holds:
\\
(1) The cell $\sigma_{i}=\vororder{s_{i}, s_{i-1}, \ldots, s_{1}} \neq \emptyset$
exists, and
\\
(2) point $x_{i+1}$ dissects $\sigma_i$ into $i+1$ cells
$\sigma_{i,1}, \ldots, \sigma_{i,i+1}$ where $\sigma_{i,1} =
\{s_{i}, \ldots ,s_{1},s_{i+1}\}$ and $\sigma_{i,j} =
\{s_{i}, \ldots, s_{j}, s_{i+1}, s_{j-1}, \ldots, s_{1}\}$, for $j>1$.
\\
We prove these invariants by induction on $i$. Consider $i=1$.
Then obviously $\vororder{s_{1}}$ exists and by adding $s_{2}$ to any
$x_{2}>x_{1}$ we get that $\sigma_{1}=\{x ~\mid~ x \leq m_{1,2}\}$ and $\sigma_{2}=\{x ~\mid ~x \geq m_{1,2}\}$.
We now assume that (1) and (2) hold for step $i-1$ and consider step $i$.
By the inductive assumption, $\sigma_{i-1} \neq \emptyset$ at step $i-1$
and according to (2), $x_{i}$ dissects $\sigma$ into $i$ segments,
one of which is  $\sigma_{1,i-1}=\sigma_{i}$, which establishes (1) for step $i$.
It is left to show that $x_{i+1}$ breaks $\sigma_{i}$ into $i+1$ segments, i.e.,
$\sigma_{i,1}, \ldots, \sigma_{i,i+1}$. Note that $\sigma_{i,j}$, for $j \in\{1, \ldots, i\}$, exists if and only if $m_{j,i+1} \in \sigma_{i}$. To see that the latter fact indeed holds, observe that
$\sigma_{i}=\{x \geq m_{i-1,i}\}$ and that
$m_{1,i+1} \leq m_{2,i+1}, \ldots, \leq m_{i,i+1}$.
Therefore it is sufficient to show that $m_{1,i+1}\in \sigma_{i}$.
Note that $m_{i,i+1}\geq m_{i-1,i}$ since $x_{i+1}>x_{i-1}$,
and that $m_{1,i+1} \geq m_{i-1,i}$ since $x_{i+1} > x_{i}+x_{i-1}-x_{1}$,
and therefore (2) holds as well.
\\
So far, we showed that step $i$ of the iterative process, adds $i$ new cells,
such that $\overrightarrow{S}_{j}\in \NCO_1$, due to $s_{i}$,
namely, $\sigma_{i-1,1}, \ldots, \sigma_{i-1,i}$.
Observe now that at each step $i >2$, there are $i-2$ newly added  cells $\vororder{S_j}$, where $S_{j} \in \NCO_1$. This follows by noting that $i-1$ among the $i$ newly added cells correspond to $\NCO_1$, and therefore after $n$ steps we end with $\Omega(n^{2})$ cells of $\NCO_1$.
This establishes part (b) of the lemma.
\QED
Exploiting the relation between
SIC-SINR diagrams and high-order Voronoi diagrams, we establish the following.
\begin{lemma}
\label{cor:nzones_rd}
$\NZones_{i}^{SIC}=O(n^{2d})$, for every $s_{i} \in S$.
\end{lemma}
\Proof
By Lemma \ref{cor:H_S_cells_vor} and Eq. (\ref{eq:zone_cancellation}),
a cancellation ordering  might correspond to a distinct cell in
$\ReceptionZone^{SIC}(\Station_1)$ only if it is in $\NCO_{1}$.
Therefore the lemma follows immediately by Lemma \ref{lem:order_ncells}.
\QED

\subsection{A Tighter Bound on the Number of Connected Components.}
\label{subsection:tight_bound}

In this section we introduce a key parameter of a wireless network \( \cA = \langle d, S, \overline{1}, \Noise, \beta, \alpha \rangle \), termed the
{\em Compactenss Parameter} of the network, defined as
$$\CompactnessParameter(\cA) ~=~ \beta^{1/\alpha}.$$
In what follows, we show that in the SIC-SINR model, this parameter
plays a key role affecting the complexity of the resulting diagram.
In particular, it follows that when $\alpha \to \infty$, both
$\CompactnessParameter(\cA) \to 1$ and the number of components gets closer
to the bounds dictated by the high order Voronoi diagram.
However, for a certain threshold value of  $\CompactnessParameter(\cA)$,
the situation is guaranteed to be much better,
as discussed later on.
Our results motivate further study of the compactness parameter,
towards better understanding of the dynamics of the SINR diagram
as a function of its compactness.

A network $\cA$ is \emph{compact} if $\CompactnessParameter(\cA) \geq 5$.
(The precise constant can in fact be slightly smaller; no attempt was made to optimize it.)
Toward the end of this subsection, we establish the following properties of compact networks.
\begin{lemma}
\label{lemma:nzones_rd}
If the network $\cA$ is compact then (a) $\NZones_{i}^{SIC}(\cA)=O(1)$, and (b) $\NZones^{SIC}(\cA)=O(n)$.
\end{lemma}
Note this this lemma implies that despite the fact that there exist
instances in $\R^{1}$ of station sets $S$ admitting $\NCO_i$ sequences
of length $\Omega(n^{2})$ (see Lemma \ref{lem:order_ncells}(b)), only a constant
number of those orderings correspond to nonempty reception regions.

It is important to understand the practical implications of $\CompactnessParameter(\cA)$. In a reasonable wireless scenario, $\alpha$ is $2$, or a small positive constant.
The threshold parameter $\beta$, however, is varying-
the higher the required rate (e.g., using higher modulation), the higher $\beta$ should be.
Thus, the above results can be interpreted as follows: the higher the rate required in the network, the less fragmented the reception region becomes. This is also a very intuitive result: under a high rate requirement, many of the small fragments of the reception zone, which require complex cancelling sequences, turn out to be empty, and only the substantial parts remain.

Hereafter, we focus on a station $s_i$, and show that $\NZones^{SIC}_i=O(1)$.
The station $s_j$ is called a {\em contributor} for station $s_i$ if its Voronoi cell $\vor(s_j)$ intersects with the reception zone of $s_i$, i.e., $\HH^{SIC}(\Station_i) \cap \vor(s_j) \neq \emptyset$.
In Lemma \ref{lem:canc_seq_rd}, we show that each contributor can contribute at most one reception cell to $\HH^{SIC}(\Station_i)$. Then, in Lemma \ref{lem:number_vors_in_siczone},
we argue that the number of contributors for station $s_i$ is $O(1)$. Combining these lemmas establishes Lemma \ref{lemma:nzones_rd}.
We begin with a general claim that holds for every metric. Consider two stations $s_i,s_j\in S$ and a point $p\in \R^{d}$.
\begin{observation}
\label{claim: dist(s',p)<comapc dist(s'',p)}
If $p\in\HH^{SIC}(s_i)$ and $\dist{s_j,p}>\dist{s_i,p}$, then
$\dist{s_j,p}\geq\CompactnessParameter(\cA)\cdot\dist{s_i,p}$. \\
\end{observation}
\Proof
If $p\in\HH^{SIC}(s_i)$ and $\dist{s_j,p}>\dist{s_i,p}$, then by Eqs. (\ref{eq:dist-order}), (\ref{eq:zone_cancellation}) and (\ref{eq:SIC_condition_zone}) there exists some ordering $\overrightarrow{S}_k$ such that $p\in\HORDER{S_k}$, $\LAST(S_k)=s_i$ and $s_j \not\in S_k$.
In other words, since
$\dist{s_j,p}>\dist{s_i,p}$ it implies that also $\Energy_{\cA}(s_j,p) <\Energy_{\cA}(s_i, p)$. Hence, assuming SINR threshold $\beta>1$, $p$ can never decode the signal of $s_j$ \emph{before} decoding the signal of $s_i$, which dominates it. This implies that
$$\left(\frac{\dist{s_j,p}}{\dist{s_i,p}}\right)^\alpha\geq\frac{\dist{s_i,p}^{-\alpha}}{\sum_{s\in S\setminus (S_k\setminus\{s_i\})}\dist{s,p}+\Noise}\geq \beta,$$
which yields the claim.
\QED
We now claim that each contributor can contribute at most one reception cell to $\HH^{SIC}(\Station_i)$.
In particular, we show the following.
Let $\overrightarrow{S}_{i}=(s_{i_1}, \ldots,s_{i_{k_1}})$ and $\overrightarrow{S}_{j}=(s_{j_1}, \ldots,s_{j_{k_2}})$,
such that $k_1 \leq k_2$ and $S_{i} \nsubseteq S_{j}$.
\begin{lemma}
\label{lem:canc_seq_rd}
If  $\HORDER{S_{i}}, \HORDER{S_{j}} \neq \emptyset$, then $s_{i_1} \neq s_{j_1}$
or $\LAST(S_{i}) \neq \LAST(S_{j})$, assuming
$\cA$ is compact.
\end{lemma}
\Proof
Assume, toward contradiction, that there exist $i$ and $j$ such that
$\HORDER{S_{i}},\HORDER{S_{j}} \neq \emptyset$ and yet
$s_{i_{1}} = s_{j_{1}}$ and $\LAST(S_{i}) = \LAST(S_{j})$.
Let $m$ be the first index such that $s_{i_{m}} \neq s_{j_{m}}$
(such $1 \leq m \leq k_1$ must exists since  $S_{i} \nsubseteq S_{j}$).
Let $S^{*}=\{s_{i_{1}}, \ldots s_{i_{m-2}}\}$ and set
$s^{*}_{1}=s_{i_{m-1}}$, $s^{*}_{2}=s_{i_{m}}$ and $s^{*}_{3}=s_{j_{m}}$.
Consider the reception zones
$X_1=\HCON(s^{*}_{1},s^{*}_{2} \mid S \setminus S^{*})$ and
$X_2=\HCON(s^{*}_{1},s^{*}_{3} \mid S \setminus S^{*})$.
Since $\HORDER{S_{i}}  \subseteq X_1$ and $\HORDER{S_{j}}  \subseteq X_2$,
it follows that $X_1, X_2 \neq \emptyset$.
Consider points $p \in X_1$ and $q \in X_2$.
For ease of notation, let $\dist{s^{*}_{1},p}=1$, $\dist{s^{*}_{2},p}=c_{2}$
and $\dist{s^{*}_{3},p}=c_{2} \cdot c_{3}$.
Since $p \in X_1$, it holds that $c_{2},c_{3} \geq \CompactnessParameter(\cA)$.
Hence, by the triangle inequality, we get that
\begin{eqnarray}
\dist{s^{*}_{1},s^{*}_{2}} &\leq& c_{2}+1,
\label{eqn:2d_cell_bound_ineq1}
\\
\dist{s^{*}_{2},s^{*}_{3}} &\geq& c_{2} \cdot c_{3}-c_{2}~.
\label{eqn:2d_cell_bound_ineq2}
\end{eqnarray}
In addition, note that by Eq. \ref{eq:dist-order}, point $q$ satisfies
$\dist{s^{*}_{3},q} < \dist{s^{*}_{2},q}$
(since $q$ cancels $s^{*}_{3}$ before $s^{*}_{2}$ and as $\beta>1$ it holds that $\Energy_{\cA}(s^{*}_{3}, q) >\Energy_{\cA}(s^{*}_{2}, q)$).
Hence by the triangle inequality we get
$\dist{s^*_3, s^*_2}\leq \dist{s^*_3,q}+\dist{s^*_2,q} \leq 2\dist{s^*_2,q}$ and combining this with Ineq. (\ref{eqn:2d_cell_bound_ineq2}), we get that
\begin{eqnarray}
\dist{s^{*}_{2},q} \geq \frac{c_{2} \cdot c_{3}-c_{2}}{2} \label{eqn:2d_cell_bound_ineq3}
\end{eqnarray}
By the triangle inequality we have that $\dist{s^*_1,q} \geq \dist{s^*_2,q}-\dist{s^*_1,s^*_2}$, hence by Ineq. (\ref{eqn:2d_cell_bound_ineq1}) and (\ref{eqn:2d_cell_bound_ineq3}) we have that
\begin{eqnarray}
\dist{s^{*}_{1},q} &\geq& \frac{c_{2} \cdot c_{3}-3c_{2}-2}{2}~. \label{eqn:2d_cell_bound_ineq4}
\end{eqnarray}
Next, note that $\dist{s^{*}_{2},q} \leq \dist{s^{*}_{1},q}+c_{2}+1$
by Ineq. (\ref{eqn:2d_cell_bound_ineq1}).
Combining this with Ineq. (\ref{eqn:2d_cell_bound_ineq4}), we get that
\begin{eqnarray*}
\frac{\dist{s^{*}_{2},q}}{\dist{s^{*}_{1},q}} &\leq&
\frac{\dist{s^*_1,q}+c_2+1}{\dist{s^*_1,q}} \leq
1+\frac{2(c_2+1)}{c_3 \cdot c_2 -3c_2-2}
< \CompactnessParameter(\cA)~,
\end{eqnarray*}
for $\CompactnessParameter(\cA) \geq 4.2$.  It therefore follows that $q \notin \HCON(s^{*}_{1}\mid S \setminus S^{*} \cup \{s^{*}_{3} \})$
and therefore also $q \notin X_2$, contradiction.
\QED

\begin{corollary}
\label{cor:canc_seq_rd}
Each contributor $s_j$ can contribute at most one reception cell to $\HH^{SIC}(\Station_i)$.
\end{corollary}
\Proof
This follows immediately by Lemma \ref{lem:canc_seq_rd}.
Assume towards contradiction that there are at least
two non empty $s_i$ reception cells $\HORDER{S_{i}},\HORDER{S_{i'}} \subseteq \HH^{SIC}(\Station_i)$ that intersect with $\vor(s_j)$.
Then, the first station in these cancellation sequences is $s_{i_1}=s_{i'_{1}}$ and the last station is $\LAST(S_{i})=\LAST(S_{i'})=s_i$, which contradicts Lemma \ref{lem:canc_seq_rd}.
\QED

We proceed by establishing a technical lemma that plays a key role in the subsequent analysis. Consider a $3$-node compact network \( \cA_{3} = \langle d, \{\Station_1, \Station_2,\Station_3\}, \Power=\overline{1}, \Noise, \beta, \alpha \rangle \) satisfying $\HH(s_1,s_2) \neq \emptyset$.
\begin{lemma}
\label{claim:compact_three_network}
(a) $\HH(s_2,s_3)=\emptyset$; \\
(b) $\HH(s_3,s_i)=\emptyset$ for each $i \in \{1,2\}$. (I.e., $\HH^{SIC}(\Station_3) \cap \vor(\Station_{i})=\emptyset$).
\end{lemma}
\Proof
Let $\cA_3' = \langle d, \{\Station_1, \Station_2,\Station_3\}, \Power,0, \beta, \alpha \rangle$ be a network similar to $\cA$ except with zero background noise. Since $\HH_{\cA_3}(s_i,s_j) \subseteq \HH_{\cA_3'}(s_i,s_j)$ for every $i,j \in \{1,2,3\}$, it suffices to prove the claims on $\cA_3'$, i.e., show that the required reception regions are empty in $\cA_3'$, which would imply the same in $\cA_3$. To avoid cumbersome notation, we continue to write $\HH(s_i,s_j)$ when referring to
$\HH_{\cA_3'}(s_i,s_j)$.
\par We begin with Part (a).
Let $p \in \HH(s_1,s_2)$. Without loss of generality assume $\dist{\Station_1,p}=1$, $\dist{\Station_2,p}=c_1$ and $\dist{\Station_3,p}=c_2 \cdot c_1$. By Claim \ref{claim: dist(s',p)<comapc dist(s'',p)}, it then follows that $c_1,c_2 \geq \CompactnessParameter(\cA)$.  Combined with the triangle inequality,
\begin{equation}
\label{eq:s_2s_3}
\dist{\Station_1,\Station_3} \geq c_1 \cdot \CompactnessParameter(\cA) -1 \mbox{~~~and~~~}
\dist{\Station_2,\Station_3}  \geq c_1 \left( \CompactnessParameter(\cA) -1\right).
\end{equation}
Assume, toward contradiction, that $\HH(s_2,s_3) \neq \emptyset$ and let $q \in \HH(s_2,s_3)$. Since $q \in \vor(s_2)$, we get that
\begin{equation}
\label{eq:trtmp}
\dist{s_2,q}\leq \dist{s_3,q}~.
\end{equation}
So by the triangle inequality, $\dist{s_3,q} \geq \dist{s_2,s_3}/2$.
Plugging in the second part of Eq. (\ref{eq:s_2s_3}) it then holds that
\begin{equation}
\label{eq:ds_3}
\dist{s_3,q}\geq c_1 \left( \CompactnessParameter(\cA) -1\right)/2.
\end{equation}
On the other hand,
\begin{eqnarray}
\nonumber
\dist{s_1,q} &\leq& \dist{s_1,p} +\dist{p,s_2}+\dist{s_2,q}
=1+c_1+\dist{s_2,q}
\\
&\leq& 1+c_1+\dist{s_3,q}
\label{eqn:s_1s_2s_3compact}
\end{eqnarray}
where last inequality holds by Eq. (\ref{eq:trtmp}). Therefore
using Eq. (\ref{eqn:s_1s_2s_3compact}) and (\ref{eq:ds_3}),
\begin{eqnarray}
\label{eqn:s_1s_2s_3compact_frac}
\frac{\dist{s_1,q}}{\dist{s_3,q}} &\leq& \frac{1+c_1+\dist{s_3,q}}{\dist{s_3,q}} \nonumber
\leq \label{eqn:s_1s_2s_3compact_frac_second}
1+\frac{2(1+c_1)}{c_1 \cdot \left( \CompactnessParameter(\cA) -1\right)}< \CompactnessParameter(\cA).
\end{eqnarray}
where the last inequality follows as $\CompactnessParameter(\cA) \geq 3$.
It then follows by Observation \ref{claim: dist(s',p)<comapc dist(s'',p)} that $q \notin \HH(s_2,s_3)$, in contradiction to the choice of $q$. Part (a) follows.\\

We now consider Part (b). Assume, toward contradiction, that $\HH(s_3,s_i) \neq \emptyset$ for some $i \in \{1,2\}$. Let $q \in \HH(s_3,s_1) \cup \HH(s_3,s_1)$. By the triangle inequality, $\dist{s_i,q}+\dist{q,s_3}\geq \dist{s_i,s_3}$ and since $q \in \vor(s_3)$, we get $\dist{s_i,q} \geq \dist{s_i,s_3}/2$ for any $i \in \{1,2\}$. Combining this with Eq. (\ref{eq:s_2s_3}), we get that
\begin{equation}
\label{eq:p2_s1_s2}
\dist{s_1,q} \geq \left(c_1 \cdot \CompactnessParameter(\cA)-1\right)/2 \mbox{~~and~~}
\dist{s_2,q} \geq c_1 \cdot \left(\CompactnessParameter(\cA)-1\right)/2.
\end{equation}
On the other hand, by the triangle inequality we get that
\begin{equation}
\label{eq:p2_s1}
\dist{s_1,q} \leq \dist{s_2,q}+c_1+1 ~~~\mbox{and}~~~
\dist{s_2,q} \leq \dist{s_1,q}+c_1+1.
\end{equation}
Combining Eq. (\ref{eq:p2_s1_s2}) and the first part of Eq. (\ref{eq:p2_s1}) we get that
\begin{eqnarray*}
\frac{\dist{s_1,q}}{\dist{s_2,q}} &\leq& \frac{\dist{s_2,q}+c_1+1}{\dist{s_2,q}} \nonumber
\leq 1+\frac{2(1+c_1)}{c_1 (\CompactnessParameter(\cA)-1)}
\\&\leq&
1+\frac{2}{\CompactnessParameter(\cA)-1}+\frac{2}{\CompactnessParameter(\cA)^2-\CompactnessParameter(\cA)}
< \CompactnessParameter(\cA).
\end{eqnarray*}
where the last inequality follows as $\CompactnessParameter(\cA)>3$. Similarly, combining Eq. (\ref{eq:p2_s1_s2}) and the second part of Eq. (\ref{eq:p2_s1}), we get
\begin{eqnarray*}
\frac{\dist{s_2,q}}{\dist{s_1,q}} &\leq& \frac{\dist{s_1,q}+c_1+1}{\dist{s_2,q}} \nonumber
\leq
1+\frac{2 \left( c_1+1\right)}{c_1 \cdot \CompactnessParameter(\cA)-1}<
\CompactnessParameter(\cA).
\end{eqnarray*}
Therefore, by Claim \ref{claim: dist(s',p)<comapc dist(s'',p)}, we have that $q \notin \HH(s_3,s_i)$ for $i \in \{1,2\}$, in contradiction to the choice of $q$. Part (b) follows.
\QED
We complete the proof by showing that every station $s_i$ has a constant number of contributors.
\begin{lemma}
\label{lem:number_vors_in_siczone}
The number of contributors for station $s_i$ is $O(1)$.
\end{lemma}
\Proof
Let $s_i$ be a contributor of station $\Station_1$ and let $\orderedSi=(s_i, s_{i_2}, \ldots, s_{i_{k-1}}, s_1) \subseteq S$ be the corresponding cancellation ordering. Let $k = |\orderedSi|$. We begin by showing that $\HH^{SIC}(\Station_1) \cap \vor(\Station_{\ell}) =\emptyset$ for every $\Station_\ell \in S \setminus \orderedSi$. Let $\widehat{S}_3=\{\Station_1, \Station_i,\Station_\ell\}$ and consider the $3$-node network \( \cA_{3} = \langle d, \widehat{S}_3, \Power=\overline{1}, 0, \beta, \alpha \rangle \).
Note that $\HH(s_i, s_1 \mid \widehat{S}_3) \neq \emptyset$ and therefore
by Lemma \ref{claim:compact_three_network}(b) we have that $\HH(s_\ell, s_1 \mid \widehat{S}_3) =\emptyset$, implying that $\HH^{SIC}_{\cA_{3}}(\Station_1) \cap \vor(\Station_{\ell} ~ \mid~ \widehat{S}_3)=\emptyset$.
Since $\HH^{SIC}(\Station_1) \subseteq  \HH^{SIC}_{\cA_{3}}(\Station_1)$
and $\vor(\Station_{\ell}) \subseteq \vor(\Station_{\ell} ~ \mid~ \widehat{S}_3)$ this will establish that  $\HH^{SIC}(\Station_1) \cap \vor(\Station_{\ell}) =\emptyset$ for every $\Station_\ell \in S \setminus S_i$.  If $k \leq 3$ then we are done (as there might be at most two contributors for $s_1$, namely, $s_i$ and $s_{i_2}$). Else, assume $k \geq 4$ and let $\ell \in \{3, \ldots, k-1\}$.  We consider a reduced network $\cA^{\ell}$ imposed on $S^{\ell}=\left(S \setminus S_i \right) \cup \{s_i,s_{i_2},s_{\ell},s_1\}$.
Assume, toward contradiction, that $\HH^{SIC}(s_1) \cap \vor(s_{\ell}) \neq \emptyset$. It then follows that also $\HH^{SIC}(s_1 \mid S^{\ell}) \cap \vor(s_{\ell}\mid S^{\ell}) \neq \emptyset$.
Note that $\HH(s_i, s_{i_2}, s_{\ell},s_1 \mid S^{\ell}) \neq \emptyset$.
To get a contradiction, we would like to show that $\HH(s_{\ell}, s_{m} \mid S^{\ell})=\emptyset$ for every $s_{m} \in S^{\ell} \setminus \{s_{\ell}\}$.
We consider three cases.  The first is when $s_{m} \in S^{\ell} \setminus \{s_1, s_{i_2},s_{\ell},s_i\}$.
Let $S_{3}=\{s_1,s_{\ell},s_{m}\}$.
Since $\HH(s_{\ell},s_1 ~\mid \widehat{S}_3) \neq \emptyset$ it follows by Lemma \ref{lem:canc_seq_rd} that $\HH(s_{\ell},s_m \mid S_3)=\emptyset$, and therefore also $\HH(s_{\ell},s_m \mid S^{\ell})=\emptyset$ (since $S_3 \subset S^{\ell}$).
The second case is when $s_{m}=s_{1}$.
Let $\widehat{S}_3'=\{s_1, s_{i_2}, s_{\ell}\}$. Then $\HH(s_{i_2}, s_{\ell} \mid S_3') \neq \emptyset$ and by Lemma \ref{claim:compact_three_network}(a), we get that
$\HH(s_{\ell},s_1 \mid S_3') =\emptyset$, implying that $\HH(s_{\ell},s_1 \mid S^{\ell})=\emptyset$ (since $S_3' \subset S^{\ell}$).  Finally, we consider the case where $s_{m} \in \{s_i,s_{i_2}\}$. Let $S_3''=\{s_i, s_{i_2}, s_{\ell}\}$. Then $\HH(s_{i}, s_{i_2} \mid S_3'') \neq \emptyset$ and by Lemma \ref{claim:compact_three_network}(b), we get that
$\HH(s_{\ell},s_i \mid S_3''),\HH(s_{\ell},s_{i_2} \mid S_3'') =\emptyset$, implying that $\HH(s_{\ell},s_i \mid S^{\ell}), \HH(s_{\ell},s_{i_2} \mid S^{\ell})=\emptyset$ (since $S_3'' \subset S^{\ell}$).\\
Overall, we get that $\HH(s_{\ell}, s_{m} \mid S^{\ell})=\emptyset$ for every $s_{m} \in S^{\ell} \setminus \{s_{\ell}\}$, concluding that $\HH^{SIC}_{\cA^{\ell}}(s_1) \cap \vor(s_{\ell} \mid S^{\ell}) =\emptyset$, in contradiction to the fact that $\HH^{SIC}(s_1) \cap \vor(s_{\ell}) \neq \emptyset$. The lemma follows.
\QED

We are now ready to conclude the proof of Lemma \ref{lemma:nzones_rd},
by combining the above two lemmas.
\Proof
By Lemma \ref{lem:canc_seq_rd}, a given Voronoi cell can contribute at most
one cell to a given $\HH^{SIC}(\Station_{i})$.
By Lemma \ref{lem:number_vors_in_siczone}, at most a constant number of Voronoi
cells can contribute to $\HH^{SIC}(\Station_{i})$.
Overall, $\HH^{SIC}(\Station_{i})$ is composed of a constant number of cells and as there are $n$ stations, overall, $\NZones^{SIC}(\cA)=O(n)$.
\QED

\section{Construction of SIC-SINR Maps}
\label{sec:Cancellation Tree}
\def\LABEL{{\cal L}}

The goal of this section is to provide an efficient scheme for constructing
$\HH^{SIC}(\Station_1)$, the reception zone under SIC for station $\Station_1$.
Recall that $\HH^{SIC}(\Station_1)$ is a collection of cells,
each corresponding to a unique cancellation ordering $\orderedSi$.
For a given network $\cA$, the reception map without SIC can be drawn by using
the characteristic polynomial of each zone $\HH(\Station_i)$.
\commabs
see Eq. (\ref{eq:reception_polynomial}).
\commabsend
In the SIC setting,
\commabs
of interference cancellation,
\commabsend
however, the characteristic polynomial of a given cell depends on the
cancellation ordering that generated it.
This is due to the fact that $\HORDER{S_{i}}$
is characterized by $|S_{i}| \leq n$ intersections of convex regions
(see Eq. (\ref{eq:unique_order_zone}))
and the characteristic polynomial of each such region is known.
The main task in drawing $\HH^{SIC}(\Station_1)$ is therefore determining the
(at most) $O(n^{2d})$ orderings of $\NCO_1$ among the collection of
a-priori $(n-1)!$ orderings (i.e., $\CO_1$).
We address this challenge by constructing the arrangement $\Arrangment(S)$ and
modifying it into a data structure that contains the information of all
$\NCO_{i}$. Towards the end of the section we establish the following.
\begin{theorem}
\label{thm:high_order_construct}
A data structure $HDS$ of size $O(n^{2d+1})$ can be constructed in time
$O(n^{2d+1})$. Using $HDS$, $\NCO_{i}$ can be computed in time $O(n^{2d+1})$.
\end{theorem}
\subsection{Algorithm Description}
\label{subsec:Algorithm Description}
We begin by providing some notation.
Associate with every point $p \in \R^{d}$ a {\em label} $\LABEL(p)$, given by
$\overrightarrow{S}^{p}=(s^{p}_{1}, \ldots, s^{p}_{n})$, a sorted array of $S$
stations such that $i < j$ if $\dist{s^{p}_{i},p} \leq \dist{s^{p}_{j},p}$.
In Observation \ref{obs:cell_arrang_order} we prove that for a cell $f$
in the arrangement $\Arrangment(S)$, all points in $f$ have the same label. Observation \ref{obs:cell_arrang_order} proves also that all cell labels
are distinct.
Hence, we associate with each cell $f \in \Arrangment(S)$
a unique label by setting $\LABEL(f)=\LABEL(p)$ for some point $p \in f$.
Let $\overrightarrow{S}^{p}_{i}=(s^{p}_{1}, \ldots, s^{p}_{k})$ be the prefix
of $\overrightarrow{S}^{p}$ such that
$\LAST(\overrightarrow{S}^{p}_{i}) = s^{p}_{k} = s_{i}$.
We then define $\LABEL_{i}(p) = \overrightarrow{S}^{p}_{i}$,
and $\LABEL_{i}(f)$ be defined accordingly.

We proceed by describing Algorithm $\HDS$.
\commabs
(Fig. \ref{$HDSConstruct$}).
\commabsend
The algorithm is composed of two steps:
\\
(1) building $\Arrangment(S)$ (see Chapter 7 of \cite{Edelsbrunner-CG}), and
\\
(2) computing the labels $\LABEL(f)$ for the cells $f \in \Arrangment(S)$.
\\
The resulting data structure $HDS$ is a labeled arrangement denoted by
$\LABEL(\Arrangment(S))$. We now describe the labeling process, given
by  Algorithm $\LabelArrangment$.
\commabs
(Fig. \ref{$LabelArrangment$}).
\commabsend
The algorithm starts from an arbitrary cell $f \in \Arrangment(S)$ and computes
$\LABEL(f)$ by ordering the distances of stations in $S$ with respect to some
arbitrary point $p \in f$. Starting from $f$, $\Arrangment(S)$ is now traversed
in a DFS fashion, where the label of a newly encountered cell $g$, $\LABEL(g)$,
is computed using the label of its parent  in the DFS tree, $\LABEL(parent(g))$.
Given that $g$ and $parent(g)$ are separated by the hyperplane $HP(s_{i},s_{j})$,
Algorithm $\LabelCell$
\commabs
(Fig. \ref{$LabelCell$})
\commabsend
sets $\LABEL(g)=\LABEL(parent(g))$ and swaps the relevant positions $s_{i}$ and
$s_{j}$ in $\LABEL(parent(g))$. Finally, Algorithm $\ExtractNCO$
\commabs
(Fig. \ref{$ExtractNCO$})
\commabsend
describes how $\NCO_{i}$ can be extracted from $HDS$.
The algorithm constructs a hash-table $HDS_i$ to maintain $\NCO_i$. To do that,
the algorithm traverses the labeled arrangement $\LABEL(\Arrangment(S))$ and
appends the truncated labels $\LABEL_{i}(f)$ to $HDS_i$.

\subsection{Analysis}
We next sketch the correctness proof of the algorithm.
We begin by showing that the labels of all points in a given cell
$f \in \Arrangment(S)$ are the same and the face labels are distinct.
\begin{observation}
\label{obs:cell_arrang_order}
The points $p_{1},p_{2} \in \R^{d}$ belong to the same face in $\Arrangment(S)$
iff $\LABEL(p_1)=\LABEL(p_2)$.
\end{observation}
\Proof
First suppose, towards contradiction, that $p_{1}$ and $p_{2}$ belong to
the same cell $f$ in $\Arrangment(S)$, and yet $\LABEL(p_1) \ne \LABEL(p_2)$,
namely, there exist stations $s_{k_1}, s_{k_2}$ such that
$\dist{s_{k_1},p_1} < \dist{s_{k_2},p_1}$ but
$\dist{s_{k_1},p_2} > \dist{s_{k_2},p_2}$.
Then the hyperplane $HP(s_{k_1},s_{k_2})$ must intersect $f$, in contradiction
to the definition of $f$.
Next, consider the reverse direction. Consider $p_{1},p_{2}$ such that
$\LABEL(p_1)=\LABEL(p_2)$, and assume to the contrary that $p_{1} \in f_{1}$
while $p_{2} \in f_{2} \neq f_{1}$. Then there must exist some
$HP(s_{k_1},s_{k_2})$ that separates $f_{1}$ and $f_{2}$. Without loss
of generality, assume $\dist{s_{k_1},p_2} >\dist{s_{k_2},p_2}$.
But it then follows that  $\dist{s_{k_1},p_1} < \dist{s_{k_2},p_1}$,
in contradiction to the fact that $\LABEL(p_1)=\LABEL(p_2)$.
\QED

To show that Algorithm $\LabelCell$ is correct,
we establish the following claims.
\begin{lemma}
\label{lem:neighboring_cell_in_arrang}
Let $f_{1},f_{2}$ be two neighboring cells in  $\Arrangment(S)$.
Then given $\LABEL(f_1)$, Algorithm $\LabelCell$ computes $\LABEL(f_2)$
in time $O(1)$.
\end{lemma}
\Proof
Assuming the points are in general position,
all $n \choose 2$ hyperplanes $\mathcal{HP}(S)$ are distinct.
Let $HP(s_{k_{1}},s_{k_{2}})$ denote the hyperplane that separates $f_{1}$ and
$f_{2}$. Then $\LABEL(f_2)$ is obtained by switching the positions of
$s_{k_{1}}$ and $s_{k_{2}}$ in $\LABEL(f_1)$ and leaving all other entries intact.
This follows by noting that if some other swap is required, for example,
if $s_{\ell_{1}}$ and $s_{\ell_{2}}$ should also switch positions in $\LABEL(f_2)$,
then necessarily $HP(s_{\ell_{1}},s_{\ell_{2}})$ separates $f_{1}$ and $f_{2}$,
in contradiction to the fact that only one hyperplane $HP(s_{k_{1}},s_{k_{2}})$
can separate two neighboring cells.
\QED

Finally we show that $HDS_{1}$ contains all $\NCO_{1}$,
proving the correctness of Algorithm $\ExtractNCO$.
\begin{lemma}
\label{lem:correctness}
$\overrightarrow{S}_{i} \in \NCO_{1}$ iff $\overrightarrow{S}_{i} \in HDS_{1}$.
\end{lemma}
\Proof
Let $\overrightarrow{S}_{i} \in \NCO_{1}$ and set $k=|\overrightarrow{S}_{i}|$.
Consider the diagram $\mathcal{V}^{\langle k \rangle}$ imposed on $\Arrangment(S)$.
Then by Claim \ref{cl:high_order_arrangment}, there exists at least one cell $f \in \Arrangment(S)$ that has non-empty intersection with $\vororder{S_{i}}$ and empty intersection with any other cell $\vororder{S_{j}} \in \mathcal{V}^{\langle k \rangle}$. It then follows that $\LABEL_{1}(f)=\overrightarrow{S}_{i}$ and the first direction is established. Now assume $\overrightarrow{S}_{i} \in HDS_{1}$. Then there exists at least one cell $f$ such that $\LABEL_{1}(f)=\overrightarrow{S}_{i}$. Let $k=|\LABEL_{1}(f)|$, and let $\vororder{S_{j}} \in \mathcal{V}^{ \langle k \rangle}$ be a cell that intersects $f$. By Claim \ref{cl:high_order_arrangment}, $\vororder{S_{j}} \neq \emptyset$ is unique and therefore $\overrightarrow{S}_{j}=\overrightarrow{S}_{i}$, implying $\vororder{S_{i}} \neq \emptyset$  and the second direction is established.
\QED

It is left to consider the construction time and memory requirements of $HDS$.
By Lemma \ref{lem:arrangments}, $\Arrangment(S)$ is constructed in $O(n^{2d})$ time and maintained in $O(n^{2d})$ space. Clearly, the labeled arrangement $\LABEL(\Arrangment(S))$ is of size $O(n^{2d+1})$, as each label $\LABEL(f)$ is of size $O(n)$. We now consider the time it takes to label $\Arrangment(S)$ (i.e., the running time of Algorithm  $\LabelArrangment$). Note that the labeling of the first cell $f \in \Arrangment(S)$ takes $O(n \log n)$ time as it involves sorting.
Subsequent labels, however, are cheaper as they are computed directly
using the label of their neighbor and the hyperplane that separates them.
Overall, the labeling requires time linear in the size of the labeled arrangement
and bounded by $O(n^{2d+1})$. Finally we consider Algorithm   $\ExtractNCO$.
The extraction of $\NCO_i$ requires one pass over the labels
of $\Arrangment(S)$ cells and therefore takes $O(n^{2d+1})$ time.
This completes the proof of Theorem \ref{thm:high_order_construct}.

\def\ALGA{
\begin{figure}
\begin{center}
\framebox{\parbox{3in}{
{\bf Algorithm  $\HDS$} ($S$)
\begin{enumerate}
\item
$HDS \gets \emptyset;$
\item
Construct $\Arrangment(S)$  (see \cite{Edelsbrunner-CG}).
\item
$HDS \gets \LabelArrangment(\Arrangment(S));$
\end{enumerate}
}}
\end{center}
\caption{HDS Construction.}
\label{$HDSConstruct$}
\end{figure}
\begin{figure}
\begin{center}
\framebox{\parbox{3in}{
{\bf Algorithm  $\LabelArrangment$} ($\Arrangment(S)$)
\begin{enumerate}
\item
Pick an arbitrary $f \in \Arrangment(S)$.
\item
Let $\overrightarrow{S}^{p}$ be the list of stations $S$ sorted in
nondecreasing order of $\dist{s_{i},p}, p \in f$.
\item
$\LABEL(f) \gets \overrightarrow{S}^{p}$;
\item
For every other cell $g \in \Arrangment(S)$, $g \neq f$,
\\
set $\LABEL(g) \gets \emptyset$;
\item
Invoke $\LabelCell(f, \Arrangment(S));$
\end{enumerate}
}}
\end{center}
\caption{Labeling an arrangement $\Arrangment(S)$.}
\label{$LabelArrangment$}
\end{figure}
\begin{figure}
\begin{center}
\framebox{\parbox{3in}{
{\bf Algorithm  $\LabelCell$} ($f, \Arrangment(S)$)

{\vskip 8pt}

For every neighbor $g$ of $f$ in $\Arrangment(S)$ do:
\begin{enumerate}
\item
If $\LABEL(g) \neq \emptyset$, then do:
\begin{itemize}
\item
$\LABEL(g) \gets \LABEL(f)$;
\item
Let $HP(s_{i},s_{j})$ be the hyperplane that separates $g$ and $f$.
\item
Swap the positions in $\LABEL(g)$ of $s_{i},s_{j}$;
\end{itemize}
\item
$\LabelCell(g,\Arrangment(S))$
\end{enumerate}
}}
\end{center}
\caption{Labeling cells of arrangement $\Arrangment(S)$}
\label{$LabelCell$}
\end{figure}
\begin{figure}
\begin{center}
\framebox{\parbox{3in}{
{\bf Algorithm  $\ExtractNCO$} ($\HDS(S), \Station_i$)
\begin{enumerate}
\item
Set hash-table $HDS_{i} \gets NULL;$
\item
For every $f \in \Arrangment(S)$,
insert $\LABEL_{i}(f)$ into $HDS_{i}$;
\end{enumerate}
}}
\end{center}
\caption{Extracting $\NCO_i$ from $HDS$.}
\label{$ExtractNCO$}
\end{figure}
} 

\section{Approximate Point Location}
\label{sec:Point_location}

In this section, we utilize the topological properties derived thus far in order to address the problem of efficiently answering \emph{point location queries} under interference cancellation.
We first briefly review the topological and computational properties of the reception zones. Eq. \eqref{eq:zone_cancellation} describes the reception zone $\HH^{SIC}(\Station_i)$ as a union of cells. By Lemmas \ref{lem:H_S_cells} and
\ref{lem:uniform_cells_convex}, all cells in this union are distinct and convex.
Moreover, Eq. \eqref{eq:unique_order_zone} describes each cell as the intersection of at most $n$ SINR reception zones with no ordered cancellation. Corollary \ref{cor:nzones_rd} bounds the number of possible cells in $\HH^{SIC}(\Station_i)$, $\tau_i^{SIC}$, by $O(n^{2d})$. For compact networks, the established bound $\tau_i^{SIC}=O(1)$ (see Section \ref{subsection:tight_bound}) is tight. 
Theorem \ref{thm:high_order_construct} then establishes the existence of a polynomial-time algorithm to compute the non-empty cancellation ordering responsible for each cell. We now show that when all these properties of the reception zone are put to use, a point location algorithm with logarithmic running time can be devised. For ease of illustration, we focus here only on the 2-dimensional case.

At this point, a few remarks are in order. First, when no cancellation is used, a station $s_i$ can be heard at point $p$ only if the signal from $s_i$ is the strongest among all transmitting stations. In the uniform power scenario, this means $s_i$ is heard at $p$ only if $p$ is in the Voronoi cell of $s_i$. As a result, one could devise a point location algorithm that for a given point $p$ returns the nearest station $s_i$ and an answer to the question whether or not $p \in \HH(\Station_i)$ (with some slack). However, when cancellation is possible, several stations can be heard at $p$ simultaneously (even for $\beta>1$). Consequently, we consider here only \emph{joint station-location} queries, that is, we wish to answer the following question: given a point $p$ in the plane and a station $\Station_i$, is $\Station_i$ heard at $p$ under \emph{some ordering} of cancellations?

Second,
it is important to
note that without offline preprocessing, $\Omega(n \log n)$ time is required to answer a single point location query. When processing a large number of queries, this might be too costly, hence the need for a tailored data structure that will facilitate $O(\log n)$ time for each query.

Toward our goal, we use a number of results and data structures derived for the SINR model with no cancellation \cite{Avin2009PODC+full}. For completeness, we include the basic concepts herein. For further details, the readers are referred to \cite{Avin2009PODC+full}.
In \cite{Avin2009PODC+full}, the authors use the following procedure: for a given reception zone $\HH(\Station_i) \subset \Reals^2$, a square grid is drawn (see Figure \ref{figure:grid_one_shape}). Then, the boundary of $\HH(\Station_i)$ is traversed, marking the grid squares that intersect with the boundary (with possibly $O(1)$ additional squares in each step). These marked grid squares form the region $\HH^?(\Station_i)$, for which no conclusive answer can be returned. The interior grid squares form the region $\HH^+(\Station_i)$, for which an affirmative answer is returned. The rest of squares form $\HH^{-}(\Station_i)$, for which a negative answer is returned. It is proved therein that since the region $\HH(\Station_i)$ is convex and fat (Lemma \ref{fc:uniform_zones}), for any given $\epsilon$, one can choose the grid granularity such that $\Area(\HH^?(\Station_i)) \leq \epsilon \cdot \Area(\HH(\Station_i))$. 

\def\FIGF{
\begin{figure*}[htb]
\centering
\subfigure[One reception zone]{
\includegraphics[scale=0.4]{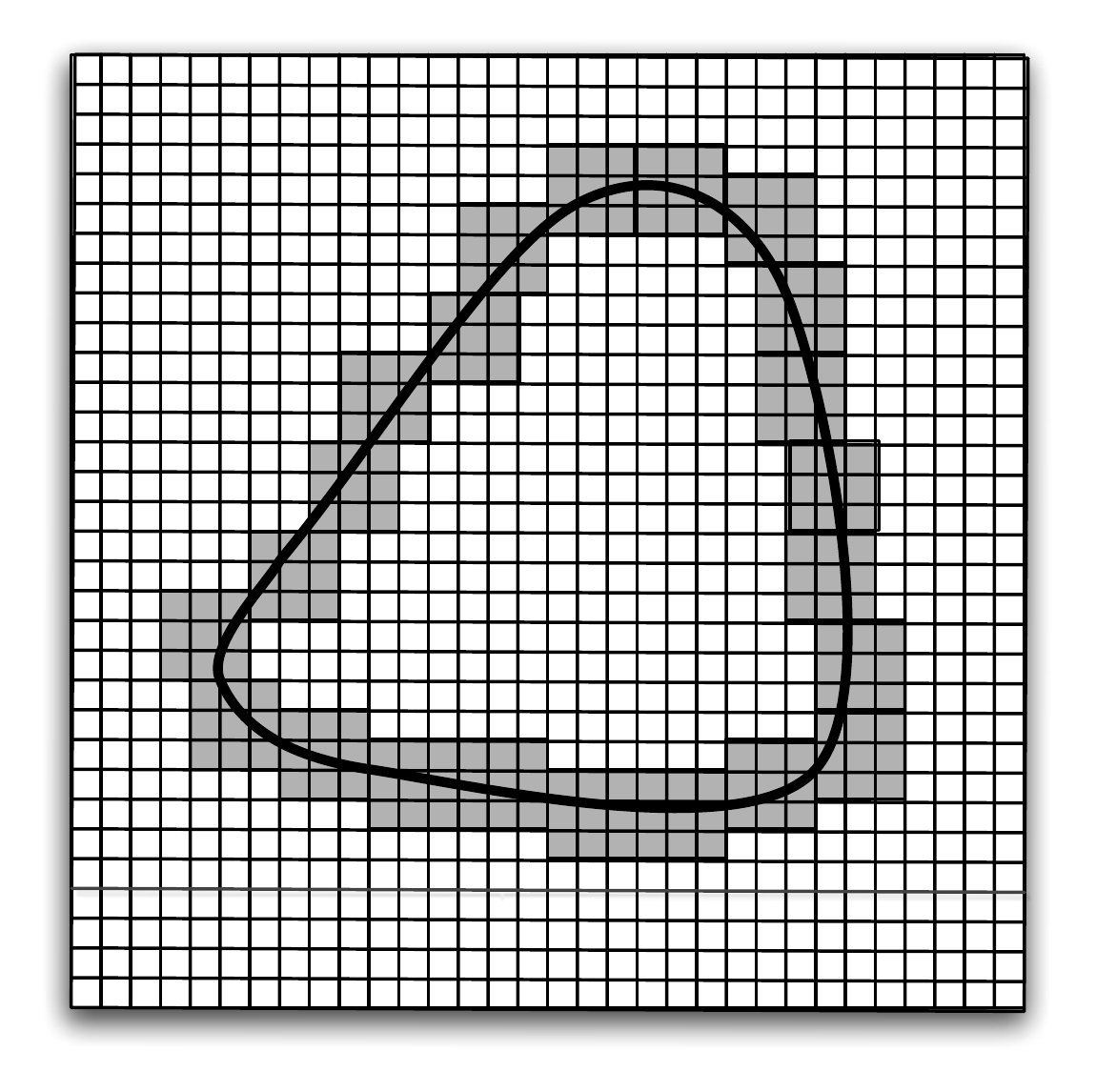}
\label{figure:grid_one_shape}
}
\subfigure[Zones intersection]{
\includegraphics[scale=0.4]{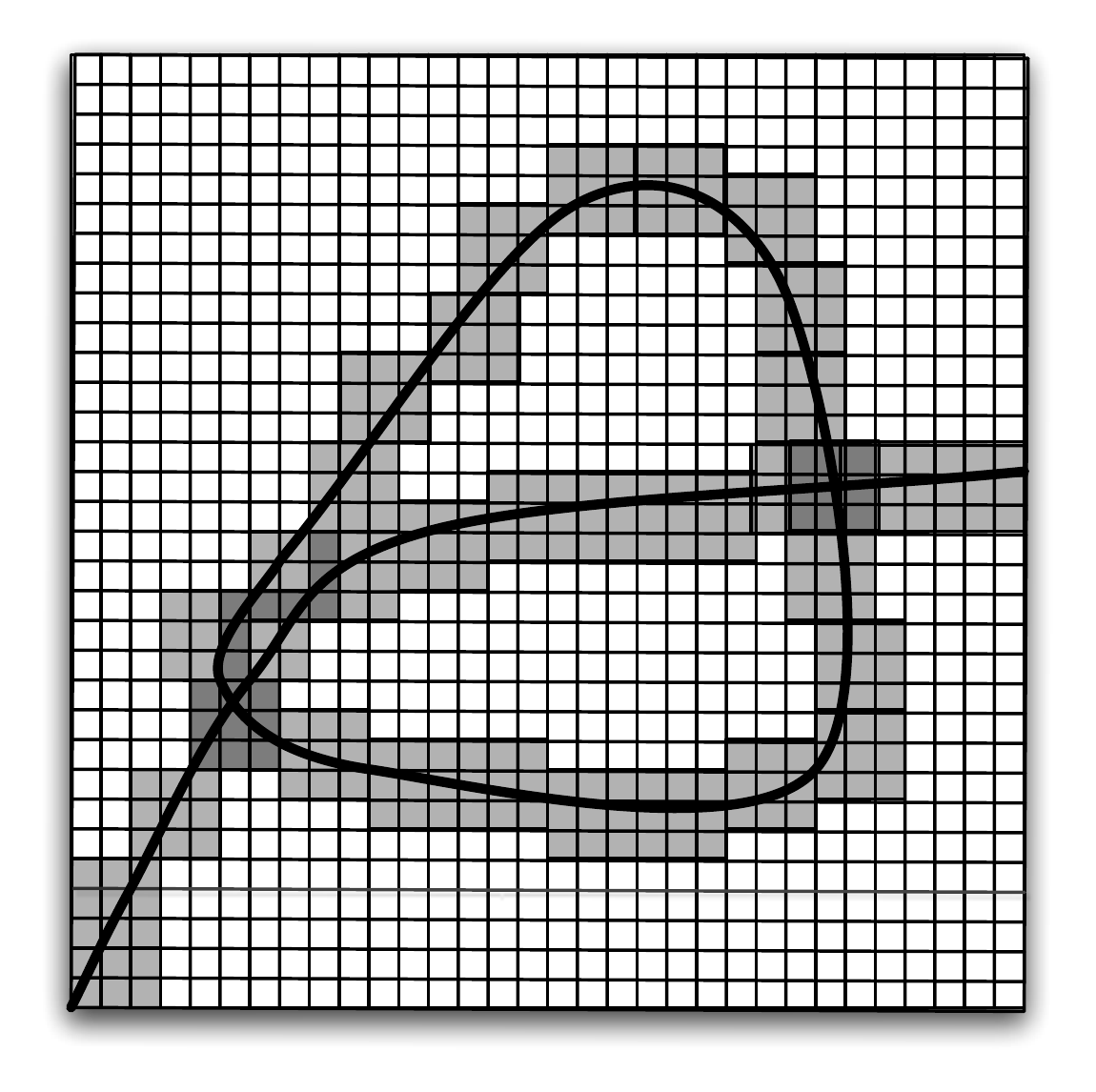}
\label{figure:grid_intersection 1}
}
\subfigure[After one iteration]{
\includegraphics[scale=0.4]{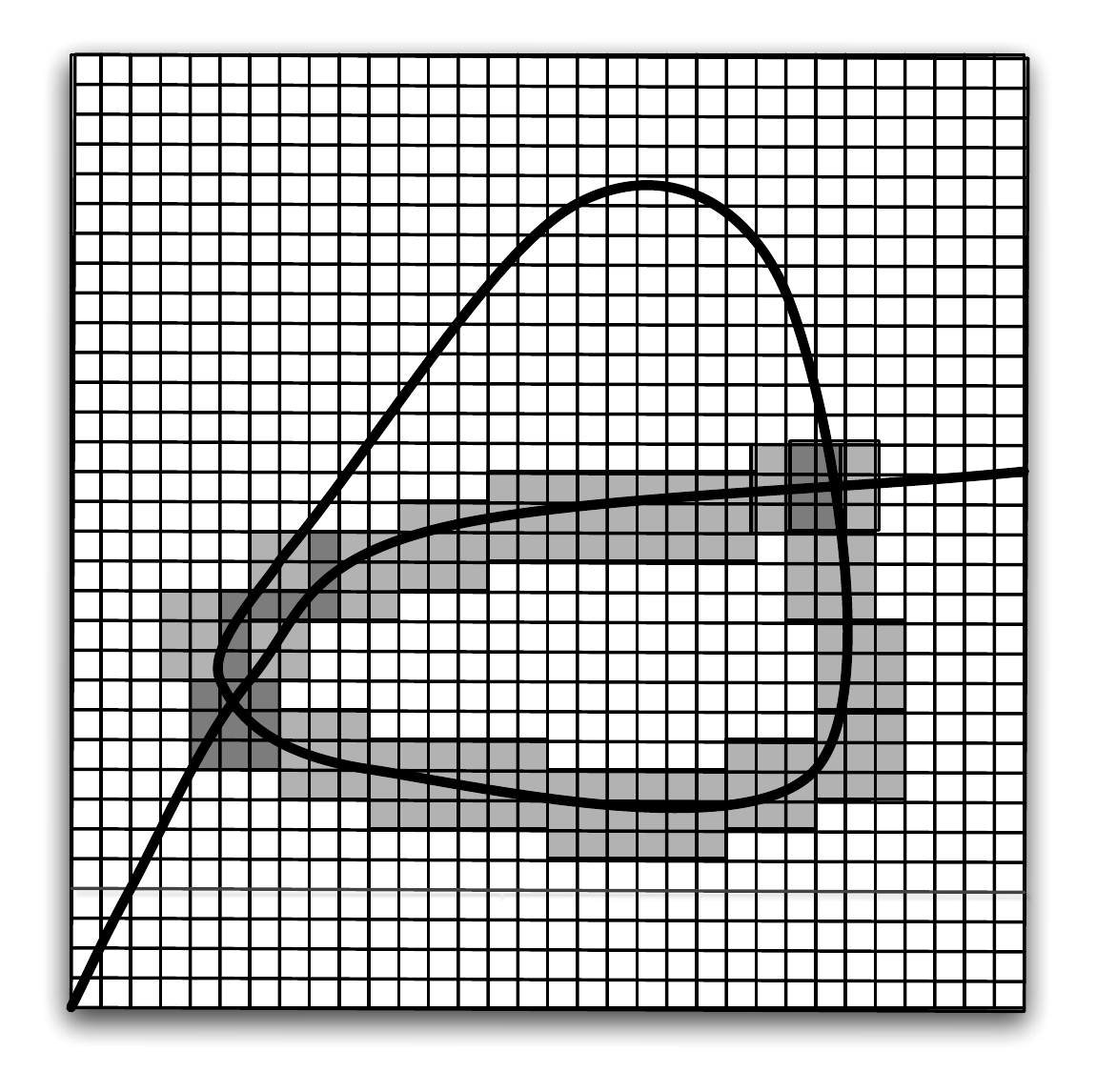}
\label{figure:grid after intersection}
}
\caption{
\sf (a) The grid structure for the representation of $\HH(\Station_i)$. The region boundary is in bold line. The undetermined squares, forming $\HH^?(\Station_i)$, are marked in gray. The inner squares form $\HH^+(\Station_i)$, while the outer form $\HH^-(\Station_i)$. (b) Adding the boundary of a second region. (c) The gray areas marking the undetermined region of the intersection. Each grayed square was grayed out in at least one of the original shapes.
}
\end{figure*}
} 
\FIGF
We now turn to our original problem. For each station $\Station_i$, we construct a data structure $\DataStructure(\Station_i)$ representing the reception zone $\HH^{SIC}(\Station_i)$, together with two binary search trees on its (now possibly slightly overlapping) cells. Using these data structures, we are able to design an algorithm answering joint station-location queries in logarithmic time. Our main result in this section is the following.
\begin{theorem}
\label{thm:point_location}
Let \( \cA = \langle d, S, \Power, \Noise, \beta, \alpha \rangle \)
for $d=2$, $\Power=\overline{1}$, $\Noise>0$, $\beta > 1$ and $\alpha=2$.
Fix a station $s_i$. A data structure $\DataStructure(\Station_i)$ of size $O(n^9\epsilon^{-1})$ is constructed in $O(n^{11}\epsilon^{-1})$ processing time. This data structure partitions the Euclidean plane into disjoint zones $\Reals^2 = \HH^{SIC,+}(\Station_i) \cup \HH^{SIC,-}(\Station_i) \cup \HH^{SIC,?}(\Station_i)$ such that
\begin{enumerate}
\item\label{item:+ area} $\HH^{SIC,+}(\Station_i) \subseteq \HH^{SIC}(\Station_i)$
\item\label{item:- area} $\HH^{SIC,-}(\Station_i) \cap \HH^{SIC}(\Station_i) = \emptyset$
\item\label{item:area of undetermined} $\Area(\HH^{SIC,?}(\Station_i)) \leq \epsilon \cdot \Area(\HH^{SIC}(\Station_i))$.
\end{enumerate}
$\DataStructure(\Station_i)$ identifies the zone to which a query point
$p\in \Reals^2$ belongs in time $O(\log n)$.
\end{theorem}

Let us start by describing the construction of the data structure
$\DataStructure(\Station_i)$.
Fix a station in the network. Without loss of generality, assume it is $\Station_1$. Represent the reception zone of $\Station_1$ as a union of cells $\HORDER{S_{i}}$ indexed by the orderings $\displaystyle{\orderedSi \in \CO_1}$, as in Eq. \eqref{eq:zone_cancellation}. Recalling that $\NCO_1 = \{ \orderedSi \subseteq S \mid \LAST(\orderedSi)=\Station_1 \mbox{~and~} \vororder{S_{i}} \neq \emptyset  \}$, in the rest of this proof, sums and unions over $\displaystyle{\orderedSi \in \CO_1}$ in the representation of $\HH^{SIC}(\Station_1)$ refer only to the distinct ordered subsets that define $\HH^{SIC}(\Station_1)$, given in $\NCO_1$, rather than to all possible subsets.

By Lemma \ref{lem:H_S_cells}, all cells in the union in the right hand side of \eqref{eq:zone_cancellation} are distinct, and each has the form
\begin{equation}
\label{eq:unique_order_zone2}
\HORDER{S_{i}} ~=~ \bigcap_{j=1}^{k} \HCON(\Station_{i_{j}} \mid
S \setminus \{\Station_{i_1}, \ldots,\Station_{i_{(j-1)}}\})~,
\end{equation}
where $k=|\orderedSi|$. $\HORDER{S_{i}}$ is thus the intersection of at most $n$ reception zones of the form $\HCON(\Station_{i_{j}} \mid S \setminus \{\Station_{i_1}, \ldots,\Station_{i_{(j-1)}}\})$, that is, SINR reception zones without interference cancellation (though with, possibly, some stations turned off). Since there are $O(n^4)$ such distinct cells, whose cancellation orders are stored in the data structure HDS, our data structure $\DataStructure(\Station_i)$
is constructed in three main steps: (1) retrieve the orderings $\displaystyle{\orderedSi \subseteq S}$ that form $\HH^{SIC}(\Station_1)$ from the data structure HDS. Now all cancellation orders in $\NCO_1$ are known. (2) Construct the data structures that represent $\HORDER{S_{i}}$, $\displaystyle{\orderedSi \in \NCO_1}$, by intersecting the required maps according to Eq. \eqref{eq:unique_order_zone2}. (3) Assemble all data structures built in step (2) in a search tree facilitating logarithmic queries.

To retrieve the orderings of cancellations that piece together $\HH^{SIC}(\Station_1)$, we traverse the data structure HDS. According to Theorem \ref{thm:high_order_construct}, $\NCO_1$ is computed and the orderings of cancellations are retrieved in $O(n^4)$ time.

We now consider the data structure $\DataStructure(\orderedSi)$ required to represent $\HORDER{S_{i}}$. Let ${\tilde\epsilon}$ be a small positive parameter, to be defined later. By \cite{Avin2009PODC+full}, for each of the reception zones in the right hand side of Eq. \eqref{eq:unique_order_zone2}, an ${\tilde\epsilon}$ approximation is achieved using a data structure of size $O({\tilde\epsilon}^{-1})$. The time required to construct such a data structure is $O(n{\tilde\epsilon}^{-1})$. Let $\DataStructure^m(\orderedSi)$, $1 \leq m \leq k$, be the data structure for the $m$th region in Eq. \eqref{eq:unique_order_zone2}. $\DataStructure^m(\orderedSi)$ partitions the space into three regions, $\Reals^2 = \HH_m^+ \cup \HH_m^-\cup \HH_m^?$ (see Figure \ref{figure:grid_one_shape}).
In particular, it is represented as a vector with $O({\tilde\epsilon}^{-1})$ entries (indexed by the $x$-axis value of the grid columns). 
Each entry stores the locations of the upper (high $y$-axis values) and lower marked squares, that is, the squares forming the boundary of the reception zone. In this way, given a point $p$, one can compute the grid square in which $p$ resides, access the data structure at the entry corresponding to the column, and based on the $y$-axis values of the upper and lower marked squares decide in $O(1)$ whether $\Station_i$ is heard at $p$, unheard, or a conclusive answer cannot be returned.
Note, however, that in order to keep all structures $\{\DataStructure^m(\orderedSi)\}_{m=1}^{k}$ on the same grid, all should be constructed according to the finest grid resolution. On the other hand, since we are interested only in the intersection, all grid columns corresponding to $x$ locations outside the region of $\DataStructure^1(\orderedSi)$ can be discarded.

To construct $\DataStructure(\orderedSi)$, we proceed as follows. We start with $\DataStructure^1(\orderedSi)$, and iterate over its ${\tilde\epsilon}^{-1}$ entries. Let $\DataStructure[l,upper]$ and $\DataStructure[l,lower]$ denote the coordinates of the undetermined upper and lower squares (respectively) in column $l$ of $\DataStructure$ (in \cite{Avin2009PODC+full}, each such region was represented by at least three squares). We say that $\DataStructure^i[l,upper] \ge \DataStructure^j[l,upper]$ if the $y$ index of the lowermost square in $\DataStructure^i[l,upper]$ is greater than or equal to the $y$ index of the lowermost square in $\DataStructure^j[l,upper]$. For each entry, we iterate over $m$, the number of cells to intersect, updating $\DataStructure^1(\orderedSi)[l,upper]$ and $\DataStructure^1(\orderedSi)[l,lower]$ in each iteration to represent the intersection of all regions at the specific entry. This is done by Algorithm  $\Intersect$, which compares the upper and lower values of the current region (describing the intersection thus far) with the ones of the new region we intersect with, and updates the region according to one of the $6$ possible intersection patters. Note that the procedure follows a simple ``hierarchy" among the three types of squares: A square that was tagged as a $'-'$ in any one of the data structures $\{\DataStructure^m(\orderedSi)\}_{m=1}^{k}$ will be tagged as such in $\DataStructure(\orderedSi)$. A square will be tagged as a $'+'$ iff it is tagged as such in all intersecting data structures. Finally, a  square tagged as a $'?'$ in $\DataStructure(\orderedSi)$ must have been tagged as such in at least one of the intersecting structures.
 An example is given in Figures \ref{figure:grid_intersection 1} and \ref{figure:grid after intersection}. In this case, a new region intersects the current one in such a way that the upper boundary of the intersection is that of the new region, and the lower boundary of the intersection is that of the current region. The comparison done in each grid column, for each new region added, takes $O(1)$ processing time. Thus, the whole process of intersecting at most $n$ given data structures requires $O(n{\tilde\epsilon}^{-1})$ processing time. The resulting data structure $\DataStructure(\orderedSi)$ representing the intersection is also of size $O({\tilde\epsilon}^{-1})$, as it is not required to be larger than the largest among the intersecting structures. Since this method requires having all data structures beforehand, these are constructed in $O(n^2{\tilde\epsilon}^{-1})$ processing time. Clearly, a data structure $\DataStructure(\Station_1)$ for representing $\HH^{SIC}(\Station_1)$ is built from the $O(n^4)$ data structures, representing all cells of $\HH^{SIC}(\Station_1)$. $\DataStructure(\Station_1)$ requires $O(n^4 \tilde\epsilon^{-1})$ memory and its construction takes
$O(n^6 {\tilde\epsilon}^{-1})$ time.
Claims \ref{item:+ area} and \ref{item:- area} of the theorem thus follow by construction.

Consider now the area of the undetermined regions in
$\DataStructure(\Station_1)$. To establish claim \ref{item:area of undetermined}
of the theorem, we wish to show that this region can be made arbitrarily small
compared to $\HH^{SIC}(\Station_1)$ with a proper choice of ${\tilde\epsilon}$.
Let $\kappa$ be the minimal distance between any two stations and define by $\HORDERQ{S_{i}}$, $\HCON^?(\Station_{i_{j}} \mid S \setminus \{\Station_{i_1}, \ldots,\Station_{i_{(j-1)}}\})$ and $\HH^?(\Station_{i_1})$ the undetermined regions in the representations of $\HORDER{S_{i}}$, $\HCON(\Station_{i_{j}} \mid S \setminus \{\Station_{i_1}, \ldots,\Station_{i_{(j-1)}}\})$ and $\HH(\Station_{i_1})$, respectively. For some constant $c_1$, we have
\begin{eqnarray*}
\Area(\HH^{SIC,?}(\Station_1)) &\leq& \sum_{\orderedSi\in\NCO_1}
\Area\left(\HORDERQ{S_{i}}\right)
~\leq~ \sum_{\orderedSi\in\NCO_1} \Area(\HH^?(\Station_{i_1}))
\\
&\leq& \sum_{\orderedSi\in\NCO_1} {\tilde\epsilon} \cdot \Area(\HH(\Station_{i_1}))
~\leq~ c_1 n^4 {\tilde\epsilon} \cdot \max_{\orderedSi\in\NCO_1}
\Area(\HH(\Station_{i_1}))
\\
&\leq& c_1 n^4 {\tilde\epsilon} \cdot \frac{n}{N\beta \kappa^2}
\Area(\HH(\Station_1))
~\leq~ \frac{c_1 n^5 {\tilde\epsilon} }{N\beta \kappa^2} \cdot
\Area(\HH^{SIC}(\Station_1))
\end{eqnarray*}
where
the first inequality holds since all cells are distinct (Lemma \ref{lem:H_S_cells});
the second is due to Eq. \eqref{eq:unique_order_zone2}
and since all cells are convex (Lemma \ref{lem:uniform_cells_convex});
the third is by the construction of the data structures $\{\DataStructure^m(\orderedSi)\}_{m=1}^k$; and
the fifth holds since the area of any reception zone with no cancellation is bounded from above by
$\frac{\pi}{N\beta}$ and from below by $\frac{\pi \kappa^2}{n}$
(\cite[Claim 15]{KLPP2011STOC}).
Choosing ${\tilde\epsilon} = \epsilon \frac{N\beta \kappa^2}{c_1 n^5}$
results in claim \ref{item:area of undetermined} of the theorem.

To achieve a query time that is logarithmic in $n$, simply arrange the $O(n^4)$ data structures representing $\HH^{SIC}(\Station_1)$ in two binary search trees, one according to right-most grid point each structure represents, and one according to the lowest grid point each structure represents. Since this procedure is merely technical, we skip the details. Given a point $p\in \Reals^2$, one can identify the data structure to which $p$ may belong in $O(\log n)$, and query the relevant data structure in $O(1)$.
%
\def\PL_COR{
Since for $d=2$, $\tau_i = O(n^4)$, we have the following corollary.
\begin{corollary}
For $d=2$, a data structure of size $O(n^9 \epsilon^{-1})$ can be built in $O(n^{11} \epsilon^{-1})$ processing time. With this data structure, the joint station-location query can be answered in $O(\log n)$ processing time, returning an undetermined answer for points in an area of at most $\epsilon$ fraction of the reception zone.
\end{corollary}
} 
\def\ALGD{
\begin{figure}[ht]
\begin{center}
\framebox{\parbox{3in}
{
{\bf Algorithm  $\Intersect$} ($\{\DataStructure^m(\orderedSi)\}_{m=1}^k$)

{\vskip 8pt}

For every $l \in {1, \ldots, \Ceil[{\tilde\epsilon}^{-1}]}$,
\\ \mbox{\hskip 20pt}
for every $m \in \{2, \ldots, m\}$:
\begin{enumerate}
\item $Upper \gets$ \\
$\min\left\{ \DataStructure^1(\orderedSi)[l,upper] ,
\DataStructure^m(\orderedSi)[l,upper]\right\}$
\item $Lower \gets \\
\max\left\{ \DataStructure^1(\orderedSi)[l,lower] ,
\DataStructure^m(\orderedSi)[l,lower]\right\}$
\item $\DataStructure^1(\orderedSi)[l,upper] \gets Upper$
\item If $Upper \geq Lower$,
\\
then $\DataStructure^1(\orderedSi)[l,lower] \gets Lower$
\item Else $\DataStructure^1(\orderedSi)[l,lower] \gets Upper$
\end{enumerate}
}}
\end{center}
\end{figure}
\vspace*{-7pt}
} 


\end{document}